\documentclass[prc,aps,groupedaddress,superscriptaddress,twocolumn,nofootinbib,showpacs,showkeys,floatfix,a4paper,10pt]{revtex4-1}

\usepackage{graphicx,colordvi,rotating}
\usepackage{multirow,color}
\usepackage{amsmath,amssymb}
\usepackage[mathscr]{euscript}
\usepackage{cancel}
\usepackage{braket}
\usepackage{cases}
\usepackage{color}
\usepackage{mathtools}
\usepackage{natbib}
\usepackage{ulem}
\usepackage{graphicx}
\usepackage{dcolumn}
\usepackage{bm}
\usepackage{natbib}
\usepackage{enumerate}
\usepackage[dvipsnames]{xcolor}
\usepackage{subcaption}
\usepackage{amsmath}
\usepackage{wrapfig}

\usepackage[utf8]{inputenc}
\usepackage{epstopdf}

\usepackage{t1enc}
\usepackage[utf8]{inputenc}

\begin{document}


\title{Spontaneous fission half-lives of actinides and super-heavy elements}

\author{J. Marin Blanco}
\email{jmarblanco@kft.umcs.lublin.pl}
\affiliation{Institute of Physics, Maria Curie--Sk\l odowska University, 20-031 Lublin, Poland}

\author{A. Dobrowolski}
\email{arturd@kft.umcs.lublin.pl}
\affiliation{Institute of Physics, Maria Curie--Sk\l odowska University, 20-031  Lublin, Poland}

\author{A. Zdeb}
\email{azdeb@kft.umcs.lublin.pl}
\affiliation{Institute of Physics, Maria Curie--Sk\l odowska University, 20-031  Lublin, Poland}

\author{J. Bartel}
\email{johann.bartel@iphc.cnrs.fr}
\affiliation{IPHC, UMR7178, University of Strasbourg, Strasbourg, France}

\date{\today}

\begin{abstract}
Spontaneous fission half-lives of actinide and super-heavy nuclei are calculated, using the
least-action integral, of the WKB tunneling probability through the barrier that appears in the
deformation landscape obtained in the macroscopic-microscopic potential-energy surface. 
This deformation-energy landscape is obtained using a Fourier shape parametrization 
with 4 deformation parameters, taking into account the nuclear elongation,
left-right asymmetry, neck formation and non-axiality degrees of freedom. 
The collective inertia tensor entering the WKB half-life expression is taken from 
the so-called irrotational flow model, whose components are scalled by an overall multiplicative factor. 
For a comparisons, we have also used the so-called phenomenological mass parameter depending only on the center-of-mass difference of the 
nascent
fission fragments. Our approach is shown to be able to reproduce empirical fission half-lives of all here considered nuclei to within 3 orders of magnitude. 


\end{abstract}


\maketitle


\section{Introduction}
Nuclear fission, as a decay mode competitive with
the emission of light particles such as neutrons or protons, light clusters like $\alpha$ particles or $\gamma$ quanta,
plays an essential role in determining the stability of heavy and super-heavy nuclei. The nuclear fission process, induced by the absorption of neutrons has been observed for the first time in 1938 by Hahn and Strassmann~\cite{HS39}. 
The theoretical explanation of this new phenomenon was given within a few weeks by Meitner and Frisch~\cite{MF39}. The authors established the basic features of the low-energy fission process, such as the energy released in this process being equal to almost 200 MeV, as well as the fact, that it results from the Coulomb repulsion of the fission fragments. In addition, it has been estimated that the number of neutrons emitted in each such fission event is larger than one, and that a chain reaction is thus possible. The spontaneous fission of uranium was discovered one and a half year later by Flerov and Petrzak~\cite{FP40}. Since these {\it early days}, a continuous interest in the theoretical description of the fission process has been observed.
  Based on the first theoretical model of a nucleus as a charged drop of liquid, fission was described as a collective motion of nucleons in which the nuclear deformation evolves from a form 
close to a sphere to an elongated shape~\cite{BW39}. Such a shape evolution is associated with the change of the nuclear deformation energy which grows with increasing deformation. When the elongation exceeds a 
certain 
critical value, the energy decreases 
again 
up to the point where the nuclear system splits into two separated fragments. In a quantum mechanical description the fission process can be understood as a tunneling through the potential energy barrier. The tunneling probability and, as a consequence, the spontaneous fission half-life strongly depends on the shape of the fission barrier, in particular its height and width.
Over the last decades, there have been numerous attempts to present a reliable model of the fission process which allows to reproduce in particular the measured spontaneous fission half-lives. What the quality of this reproduction is concerned, one
has to keep in mind, however, that already a very small
change of the barrier, in particular its height, will lead
to a substantial change in the fission half-life.
Among the best known of these early attempts, within which the
global systematics of spontaneous fission half-lives has
been reproduced, the semi-empirical formula proposed in 1955 by W.J.\ Świątecki would come immediately to the mind~\cite{S55}. The main idea of this approach comes from the observation of the strong correlation between the logarithm of the spontaneous fission half-lives and the ground state microscopic corrections due to shell effects and pairing correlations. Later on, this concept has been applied to up-to-date experimental data ~\cite{KP15,KP22} within a modern version of the liquid-drop model, which is now known as the Lublin-Strasbourg Drop (LSD)~\cite{PD03}. There are also various theoretical approaches often based on mathematically quite advanced methods \cite{BL80,SP07,VW66,CD96,RB97,baran81,smol97,staszczakPhysRevC.87.024320,wardaPhysRevC.86.014322}. There have also been several attempts to apply fully microscopic, self-consistent methods in order to reproduce spontaneous fission observables \cite{bsn11,smn}, yet the accuracy in the reproduction of the experimental data can still not be considered as being fully satisfactory. To
obtain a better agreement with the experiment, one may
consider pairing as a dynamical degree of freedom (see Refs. \cite{robpair,robpair2}) and/or use improved approaches for the collective inertia \cite{robmass}. Such approaches turn out, however, to be numerically very costly. 
In general, spontaneous fission half-life calculations require not only an assessment of the collective potential energy surface (evaluated in a purely microscopical approach or, as will be done is what follows, within the macroscopic-microscopic model), but also of the collective inertia tensor. 
%
Commonly, the latter is obtained within the Adiabatic Time-Dependent Hartree-Fock-Bogoliubov (ATDHFB) \cite{atdhfb1,atdhfb2}, the Generator Coordinate Method (GCM) with the generalized Gaussian Overlap Approximation (GOA)~\cite{GPB85,apair} or evaluated within the so-called "cranking" approximation~\cite{NTS69,BDJPSW}. In the present work the irrotational-flow approach of Ref.\ \cite{DSN76} (see also \cite{BN19}) will be used to evaluate the inertia tensor.

The present manuscript will be entirely devoted to
present spontaneous fission half-lives, obtained within
that approach, and their comparison with experimental
data.

Using
the tunneling model of the WKB approximation for a
multidimensional potential-energy barrier \cite{WKB, WKB1, WKB2}, we
will analyze the half-lives for this process for selected
even-even actinide and super-heavy nuclei from $Z$ = 90
to 110. The calculations will in particular concern the
isotopes of the following actinide isotopic chains: Th,
U, Pu, Cm, Cf, Fm, No, as well as for the superheavy
elements Rf, Sg, Hs, and Ds. The obtained results are
compared with the available experimental data. Since
this comparison turns out to be quite satisfactory, we
will also make predictions for half-lives of nuclei where
the measurements have not yet been performed.

In section II the theoretical framework of our approach will be presented with the main ingredients which are the parametrization of the nuclear shape on one side and, on the other side,
the model used to describe the energy of the nuclear system as function of the chosen deformation, which in our present study is the macroscopic-microscopic approach together with the Lublin-Strasbourg Drop model with the Strutinsky shell correction
and a seniority force BCS pairing treatment.
Section III will explain how the spontaneous-fission half-life can be evaluated in a WKB-type model, based on the least-action path, before we present in section IV our results for such half-lives for some actinide and super-heavy nuclei. Section V finally draws some conclusions and gives an outlook on further studies which can be carried out in our approach.

\section{Theoretical framework \label{THEORY}}

To be able to describe very heavy nuclei and their de-excitation through fission or particle emission, a study of the evolution of their 
energy with deformation is mandatory. We will therefore investigate in what follows the two main ingredients required for such an investigation of what is commonly called the deformation energy of the nucleus, namely the parametrization of the nuclear shape up to very large deformations as they may occur in the fission process, and a model capable to give a reliable description of the nuclear energy at a given deformation.

\subsection{Nuclear shape parametrization}

The description of the huge variety of shapes encountered all across the nuclear chart when going from oblate deformations as they appear in the transition region, generated by the progressive filling of the $pf$ shell, to prolate shapes as found in the rare-earth region and in actinide nuclei necessitates a sufficiently rich and flexible nuclear shape parametrization. This demand is even tightened if one requires to describe the typically very elongated and often necked-in shapes as they are encountered in the fission process. To model the physical reality (as far as that could be identified) as faithfully as possible, it is obviously required to involve a large number of deformation parameters,  depicting the involved degrees of freedom, characterized e.g.\ by the multipole moments of the nuclear shape. For a numerical treatment, on the other hand, a very large number of deformation parameter would be prohibitive. It is thus the demanding task for the nuclear physicist to identify the essential degrees of freedom of a nuclear shape and to bring these into an analytical form. A very large number of shape parametrizations have been proposed (see Ref.\ \cite{HM88} for an extensive review) and are currently used to investigate all kind of nuclear properties. One of the most widely used (see e.g.\ \cite{Ko1,Ko2,Ko3,Ko4}) 
such parametrization 
is the one Lord Rayleigh proposed already towards the end of the 19th century \cite{Ra79}. Among other more recent shape parametrizations which have been used to describe the fission process, one should mention the {\it quadratic surfaces of revolution} (QSR) \cite{Ni69} of Ray Nix, the Cassini ovals \cite{Pa71,PR08} of Pashkevich, the famous Funny-Hills parametrization \cite{FH72} 
of the Copenhagen group 
and its improved version \cite{PB06}, as well as the expansion of the nuclear surface in Legendre polynomials \cite{TKS80} of Trentalange, Koonin and Sierk. While the Rayleigh shapes were defined through the radius vector of any surface point in spherical coordinates $r^{}_s(\theta,\varphi)$, an approach which is certainly well adapted to the description of nuclear shapes reasonably close to a sphere, it became rapidly clear that for the description of rather elongated shapes, as they are encountered in the fission process, a parametrization that defines a surface point in cylindrical coordinates in the form $\rho^{}_s(z,\varphi)$,  
as this has been done in the Funny-Hills parametrization \cite{FH72},
is much better suited. This is e.g.\ demonstrated by the fact that if one is interested in the description of the fission process and in particular in fission barrier heights, the Rayleigh parametrization fails, or rather 
converges very slowly 
as has been demonstrated in Ref.\ \cite{DPB07}.\\

As it has already been mentioned above, the description of nuclear shapes as they appear along the way from the nuclear ground state to the pre-scission configurations is obviously not a trivial task. 
For practical reasons, 
that
description should contain as few deformation parameters of the nucleus as possible and, at the same time, reproduce at least, major classes of its shape occurring on its path to fission. Such shapes should comprise, among others, axially-symmetric and asymmetric
deformations, 
elongated forms, characterised in addition by a left-right symmetry or asymmetry, 
and the posible presence of a neck forming between the two nascent fission fragments.

To make an expansion, in cylindrical coordinates $(\rho,z,\varphi)$ of the distance $\rho_s(z)$ of any surface point to the symmetry z-axis goes back to the seminal work of Ref. \cite{FH72}  and has been developped further in \cite{TKS80} and \cite{PB06}, but to make an expansion of $\rho^2_s$ as function of the cylindrical $z$ coordinate in a Fourier series as
\begin{equation}
    {\displaystyle \frac{\rho^2_{s}(u)}{R_{0}^2} \!=\! \sum_{n=1}^{\infty}\!
      \left[a_{2n} \cos\left(\!\frac{2n-1}{2} \pi u \!\right) 
        + a_{2n+1} \sin\left(\!\frac{2n}{2} \pi u \!\right) \right]},
\label{Four}\end{equation}
%
has been presented for the first time in Ref. \cite{Pom15}. 
Here $R_{0}$ is the radius of the spherical nucleus having the same volume, while $2\,z_{0}$ is the length of the nuclear shape along the 
symmetry 
$z-$axis.
The dimensionless variable $u=(z-z_{sh})/z_0$ that appears in (\ref{Four}) contains a parameter $z_{sh}$ given by:
\begin{equation}
    {\displaystyle{z_{sh}} = \frac{3z_{0}^2}{2 \pi R_{0}} \sum_n (-1)^n\, \frac{a_{2n+1}}{n}},
\label{zsh}\end{equation}
which ensures that the center-of-mass of the shape is always located at the origin of the coordinate system.

There are of course very many parametrizations to describe the shape of a deformed nucleus. When one is interested in rather elongated shapes as they occur in the fission process it is certainly beneficial to use a parametrization defined in cylindrical rather than in spherical coordinates, as explained above. That is also why the so-called {\it Funny-Hills} shapes of Ref.\ \cite{FH72} had such a enormous succes when dealing with the fission process. The main advantage of the Fourier shapes consists in the fact that the expansion of Eq.\ (\ref{Four}) converges very rapidly and that its convergence can be tested easily by carrying the expansion to higher orders, thus including higher multipoles, which was not possible for the Funny-Hills parametrization \cite{FH72} or its extension  \cite{PB06}.
%
%

Since our Fourier parametrization constitutes, when carried to infinite order, a complete orthogonal series, it is clear that for a given underlying model (like our macro-micro approach) the quality of the parametrization will crutially depend on the convergence of the expansion, i.e. on the number of required expansion coefficients. It turns out, however, that with a very limited number (of the order of 2-3) of Fourier coefficients, which are going to be our deformation parameters, one is able to describe the nuclear energy along the fission path with a quite good accuracy of the order of half an MeV as compared to the case when higher order deformation parameters are taken into account as has been shown in Ref.~\cite{kostr_cpc21}. It also turns out that these higher-order terms will play any noticible role only for very large elongations and mass-asymmetry deformations ($q_2\approx 2.0$ and $q_3 > 0.15$), while they do not exceed a very small fraction of an MeV in the vicinity of the ground state.

The above parametrization 
(\ref{Four}) is obviously limited to axially symmetric shapes. Shapes breaking axial symmetry can, however, 
be easily taken into account by assuming that the cross section perpendicular to the symmetry $z-$axis is always of the form of an ellipse with half axes $a$ und $b$ (see Fig.\ 1), such that $a \, b = \rho_s^2(z)$ which ensures volume conservation. One then defines a non-axiality parameter: 
\begin{equation}
    \eta = \frac{b-a}{a+b}.
\label{eta}\end{equation}
which is the relative difference of the half axes $a$ and $b$. Assuming that this parameter stays the same all across the nuclear shape, the profile function of the nucleus can then be written in the general case of an axially-asymmetric shape as \cite{BPN17,SPN17}:
\begin{equation}
  \varrho_s^2(z,\varphi) = \rho_s^2(z) \, f_\eta(\varphi),
\label{rho2s}\end{equation}
where 
\begin{equation}
  f_\eta(\varphi) = \frac{1-\eta^2}{1 + \eta^2 + 2 \, \eta \cos(2 \varphi)} \,.
\label{feta}\end{equation}
\begin{figure}[!h]
  \includegraphics [scale=0.36] {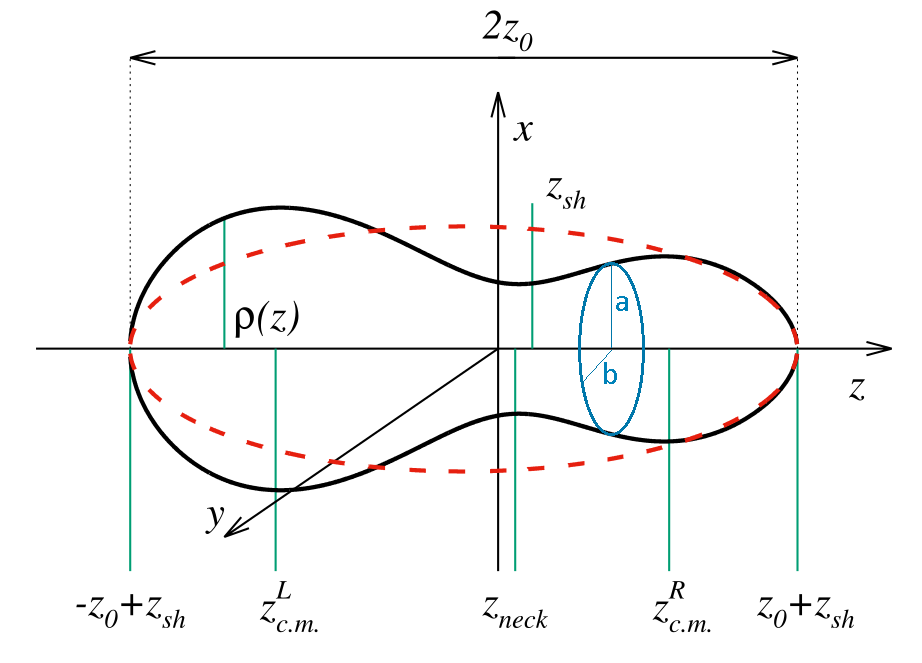}
\caption{Schematic visualization of the parameters entering the definition of the profile function defined through Eqs.~(\ref{Four}) $-$ (\ref{feta}) .}
\label{fig1}\end{figure}
In order to relate the Fourier coefficients $a_{\nu}$, which are our original deformation coordinates, to some more physical deformation parameters and make them vanish, at the same time, for a spherical shape, we introduce new collective coordinates $q_\nu$ \cite{Pom15} through
\begin{equation}
\begin{split}
  q_{2}&=a_{2}^{(0)}/a_{2}-a_{2}/a_{2}^{(0)},\\
  q_{3}&=a_{3}, \\
  q_{4}&=a_{4}+\sqrt{(q_{2}/9)^{2}+(a_{4}^{(0)})^{2}}, \\
  q_{5}&= a_5 - (q_2 - 2)\,\frac{a_3}{10}, \\
  q_{6}&=a_{6} - \sqrt{(\frac{a_2}{100})^2 + (a_6^{(0)})^2} , 
\end{split}
    \label{a_to_q}
\end{equation}
where the $a_{2n}^{(0)}$ defined by
\begin{equation}
    a_{2n}^{(0)}=(-1)^{n-1}32/[\pi(2n-1)]^{3}
\label{transf}\end{equation}
correspond to the values of the $a_{2n}$ 
for a sphere. In what follows, we will limit ourselves to only four deformation parameters $(q_2,q_3,q_4,\eta)$, where the parameter $q_2$ determines the elongation of the shape and therefore stands for the quadrupole degree of freedom, $q_3$ for the octupole deformation and thus for the left-right asymmetry and $q_4$ for the hexadecapole deformation and would be responsible for the possible formation of a neck region. Higher order terms would then define higher-order multipole moments.

\subsection{The macroscopic-microscopic approach}

Having defined above an analytical form of the parametrization of the nuclear shape that is rapidly convergent, as 
already mentioned and as this has been shown e.g.\ in Ref.\ \cite{SPN17}, we shall now present the macroscopic-microscopic approach which will allow us to evaluate the nuclear energy for any deformation that can possibly be defined through the above shape 
parametrization. 
This macroscopic-microscopic approach relies on a parametrization of the average, liquid-drop type energy in the spirit of the Bethe-Weizs\"acker mass formula \cite{We35,Be36}. The liquid-drop type approach that we are using in what follows is what is known as the Lublin-Strasbourg Drop (LSD) \cite{PD03}, which has the particularity to contain 
in the leptodermous expansion a curvature $A^{1/3}$ term 
and a 
congruence energy term \cite{Cong,MNM95}. 
The total nuclear energy is then given by: 
%
\begin{equation}
\begin{split}
E_{LSD}=&b_{vol}(1-k_{vol}I^2)A-\\
&b_{surf}(1-k_{surf}I^2)A^{2/3}B_{surf}(\{q_i\})\\
&-b_{cur}(1-k_{cur}I^2)A^{1/3}B_{cur}(\{q_i\})\\
&-\frac{3}{5}e^2\frac{Z^2}{r_0^{ch}A^{1/3}}B_{Coul}(\{q_i\})+C_4\frac{Z^2}{A}\\
&-10\,exp(-4.2|I|) .
\end{split}
\label{ELSD}
\end{equation}
Taking into account, in addition, microscopic energy corrections, taken from Ref.\ \cite{MNM95}, the thus obtained total nuclear energy, 
which had been fitted to reproduce in the best possible way the ground-state masses of the 2766 isotopes with $ N\ge 8$ and $Z\ge 8$ known at that time (2003), has been quite successful. It has, indeed, been shown that it does not only yield an excellent description of nuclear masses 
(with an r.m.s.\ deviation of 0.70 MeV from the experimental data),
but that it is also able to reproduce experimentally determined fission-barrier heights with a very good accuracy
\cite{Pom13} .
The coefficients of this leptodermous expansion are given in the Table 1.
%
%

%
%
\begin{table}[!ht]
    \centering
    \caption{Values of the parameters of the Lublin-Strasbourg Drop Model.}
    \begin{tabular}{|l|l|}
    \hline
        \textbf{$b_{vol}$} & 15.4920 MeV \\ \hline
        \textbf{$b_{surf}$} & 16.9707 MeV \\ \hline
        \textbf{$b_{cur}$} &  3.8602 MeV \\ \hline
        \textbf{$k_{vol}$} & 1.8601 \\ \hline
        \textbf{$k_{surf}$} & 2.2038 \\ \hline
        \textbf{$k_{cur}$} & -2.3764 \\ \hline
        \textbf{$C_4$} & 0.9181 MeV \\ \hline
        \textbf{$r_0$} & 1.21725 fm \\ \hline
    \end{tabular}
    \label{tab1}
\end{table}


%

\subsection{Shell and pairing corrections \label{THEORY}}

Observing that the 
average 
nuclear energy can be approximated to some reasonable extent by a macroscopic mass formula, like the one of Weizs\"acker and Bethe \cite{We35,Be36}, but that there exist quantum effects in such a microscopic system, which are often responsible for the essential physical phenomena, like the structure of the nuclear ground state, a description of the influence of these quantum effects, associated with the existence of the shell structure in nuclei, was introduced by Myers and Swiatecki in 1966 in terms of energy corrections \cite{MS66} to the smooth liquid-drop energy given in our case by Eq.\ (\ref{ELSD}).
\\[ -2.0ex]

An important contribution was then made by Strutinsky in 1968 \cite{Str66,Str67,Str68} who proposed an efficient and fast method for evaluating the total energy of a nucleus by a smoothing procedure of the single-particle spectrum which at the same time 
takes into account in some approximate way 
the influence of the energy levels lying in the continuum.
%
%
The average nuclear energy obtained in such a way can then be subtracted from the sum of the single-particle levels, to yield the so-called Strutinsky shell-correction energy 
\begin{equation}
   \delta E_{shell} = \sum_\nu \left[n_\nu - \tilde{n}_\nu \frac{}{}\!\right] e_\nu
\label{ESH}\end{equation}
where $n_\nu$ is a Heavyside step function, with values 0 or 1 depending on whether $e_\nu$ is located above or below the Fermi energy and $\tilde{n}_\nu$ is obtained by a Strutinsky smoothing procedure \cite{BP73}. 
The main advantage of the Strutinsky method, is that it can be applied to an arbitrary spectrum of single-particle states. 
\\[ -2.0ex]

Another microscopic correction to the total energy of the nucleus has its origin in the pairing correlations which exist 
in a BCS-type approach and for the heavy nuclei in our study only 
between nucleons of the same type (protons or neutrons). These pairing correlations cause nuclei having an even number of protons or neutrons to be more bound. This pairing interaction is described here by means of the superconducting approach proposed initially for the correlations between electrons by Bardeen, Cooper and Schrieffer \cite{BCSpa} in the framework 
of solid state physics. 
In order to obtain a many-body solution which is an eigenstate of the particle number operator, an approximate projection of the BCS wave functions onto good particle number is carried out in our approach using the Generator Coordinate Method (GCM) with the Gaussian Overlap Approximation (GOA), as presented e.g.\ in Ref.\ \cite{GCM+GOA}.
Let us recall that, in general, the $n^{th}$ GCM many-body state $|\Psi_n (X)\rangle$ is constructed as a function of the single-particle variables $X$ as
\begin{equation}
  |\Psi_n (X)\rangle = \int dq\,f_n(q)|X;q\rangle,
\label{genfun}\end{equation}
where $f_n(q)$ is called a weight function and $|X;q\rangle$ is a generator function (of HF or HFB eigensolutions or BCS many-body solutions) which depends on the single-particle variables $X$ and parametrically on a certain set of collective variables $\{q\}$ which can be taken simply as the nuclear deformation parameters or other degrees of freedom describing nuclear collective motions. To determine the weights $f_n(q)$, one assumes the existence of stationary solutions $\varepsilon_n$ of a many-body Hamiltonian $\hat H_{mb}$, with respect to variations $\delta f_n(q)$
\begin{equation}
\langle \Psi_n (X)|\hat H_{mb}|\Psi_n (X)\rangle\approx
\langle \Phi_n(q) |\mathcal{\hat H}_{coll}|\Phi_n(q)\rangle=\varepsilon_n.
\label{GCM_cond}\end{equation}
Such a prescription represents an approximate way of mapping the single-particle fermionic space onto a collective one, spanned by collective wave functions $|\Phi_n(q)\rangle$. For this purpose, one assumes that the generator coordinates are continuous and the overlap of generating functions $|X;q\rangle$ has the form of a multidimensional Gaussian function or may be transformed into a Gaussian shape.
Let us now choose a generator function $|X;\phi,q\rangle$ of the form
\begin{equation}
   |X;\phi,q\rangle=e^{i\phi\mathcal{\hat N}}|X;q\rangle_{{\rm BCS}},
\label{genf_BCS}\end{equation}
where $\phi$ is the so called gauge angle and $\{q\}$ is the set of our collective deformation parameters $(\eta,q_2,q_3,q_4)$. 
The hermitian operator $\mathcal{\hat N}$ describes the fluctuations of the particle number
\begin{equation}
\mathcal{\hat N}=-i\frac{\partial}{\partial\phi} \equiv \hat N -\, {}_{{\rm BCS}}\langle X;q|\hat N |X;q\rangle_{{\rm BCS}}.
\label{N_operator}\end{equation}
With the above assumptions, the 
generator function  
entering Eq.~(\ref{genfun})
may be rewritten to the form
\begin{equation}
|X;q\rangle_m =\int\limits_{0}^{2\pi}\! d\phi \; e^{i(N+m)\phi}[e^{-i{\hat N}\phi}
|X;q\rangle_{{\rm BCS}}]
\label{g_function}\end{equation}
with $m=N-\langle \hat N \rangle=0,\pm 2, \pm 4,...$ corresponding to a quantum number of rotation in the gauge space. For $m=0$ we get the prescription for the typical particle-number projected generator function of the ground state, where no quasi-particle pair is excited. The only effect of the particle number projection is then given by the appearance of a zero-point energy correction $\epsilon_{0}$ given as 
\begin{equation}
\epsilon_{0}=\frac{\sum\limits_{\nu>0} \left[(e_{\nu}-\lambda)(u_{\nu}^2-v_{\nu}^2)+2\Delta u_{\nu}v_{\nu}+G\,v_{\nu}^4 \right]/E_{\nu}^{2}}{\sum\limits_{\nu>0}E_{\nu}^{-2}},
\label{e0}
\end{equation}
which subtracted from the BCS ground-state energy without the projection effects, leads to a deeper ground-state energy at a slightly higher value of the pairing gap $\Delta$ as compared to the corresponding value in the original BCS approach without projection.
In Eq.\ (\ref{e0}), $E_{\nu}=\sqrt{(e_{\nu}-\lambda)^{2}+\Delta^{2}}$ is the quasi-particle energy
while $\lambda$ and $G$ are respectively the BCS Fermi energy and the constant pairing strength.
Let us recall that these equations need to be defined independently for protons and neutrons.
The summations in the above equations runs over the single-particle states inside what is called a pairing window  of energy width $2 \Omega$ around the Fermi energy ($\lambda - \Omega < e_{\nu} < \lambda + \Omega$). Since the pairing interaction takes place between a pair of particles in time-reversed states and since these have precisely the same energy, this summation runs only over states with one fixed orientation of the total angular momentum (let us call these $k\!>\!0$), excluding their time-reversed ($k\!<\!0$) partners.
In the above equations $v_{\nu}^2$ is the occupation probability of the single-particle state of energy $e_{\nu}$ while $u_{\nu}^2$ denotes the probability that this state is unoccupied. Obviously, $v_{\nu}^2+u_{\nu}^2=1$.
The single-particle energies $e_{\nu}$ are the eigenvalues of a mean-field Hamiltonian 
with a mean-field potential 
chosen in a well adapted way to describe the nucleus under study at the chosen deformation. In this work this is generally done by folding the deformed shape, generated in our case by the Fourier expansion described in section II.A, with a Yukawa-folded single-particle potential as explained e.g. in Refs.\ \cite{Yuk1,Yuk2}.

The energy correction generated by the pairing correlations is in general defined as 
the difference between 
the nuclear energy, obtained in the above projected $BCS$ approach  and the sum of the single particle energies up to the last occupied level 
\begin{equation}
\delta E_{pair}=E_{{\rm BCS}} - \sum_{\nu} e_{\nu}-{\tilde E_{pair}},
\label{Epaircor}\end{equation}
where $\tilde E_{pair}$ is the so-called average pairing energy which is not included in the liquid drop formula.
The ground state energy of the nucleus in such an approximation can then be written as
\begin{equation}
E_{{\rm BCS}}=2\sum\limits_{\nu>0}e_{\nu} v_{\nu}^2 - G(\sum\limits_{\nu>0} u_{\nu} v_{\nu})^{2}
 - G\sum\limits_{\nu>0} v_{\nu}^4-\epsilon_{0} \,.
\end{equation}
The average pairing energy, projected onto good particle number, then writes as
\begin{equation}
\begin{split}
\tilde E_{pair}=&-\frac{1}{2}\tilde g\,\tilde\Delta^2+
             \frac{1}{2}\tilde g\,\tilde\Delta\,\arctan\left({\frac{\Omega}{\tilde\Delta}}\right)-
             \log\left({\frac{\Omega}{\tilde\Delta}}\right)\,\tilde\Delta\\
             &+\frac{3}{4}G\,\frac{\Omega/\tilde\Delta}{1+\left(\frac{\Omega}{\tilde\Delta}\right)^2}/\arctan\left({\frac{\Omega}{\tilde\Delta}}\right)-
             \frac{1}{4}G,
\end{split}
\label{epair1}
\end{equation}
where $\tilde g$ is the average density of single-particle levels in the $2\Omega$ energy window whereas $\tilde\Delta$ denotes the average pairing gap corresponding to a given strength $G$ of pairing interaction \cite{pom06}
\begin{equation}
\tilde\Delta=2\Omega\, e^{-1/(G\tilde g)}.
\end{equation}
In all above considerations one admits a pairing energy window of width $2\Omega$, containing $2 \sqrt{15 N_q}\,\,\, (N_q\!=\!{\rm\, Z\, or\, N})$ single-particle levels around the Fermi level \cite{NTS69}.

\subsection {Fitting the pairing strength}

To be able to carry out the calculations in the above described model with pairing correlations acting inside a pairing window of width $2 \Omega$ around the Fermi energy, one has to adjust the pairing strengths $G$ and through that, the pairing gaps $\Delta(G)$ for protons and neutrons. The latter have to reproduce as accurately as possible the experimental proton and neutron pairing gaps $\Delta_q^{(exp)}$ calculated out of 
measured mass excesses of neighbouring odd-even heavy and super-heavy nuclei.
\\[ -2.0ex]

\begin{figure*}[ht!]
      \hskip -0.5cm
      \includegraphics[scale=0.172]{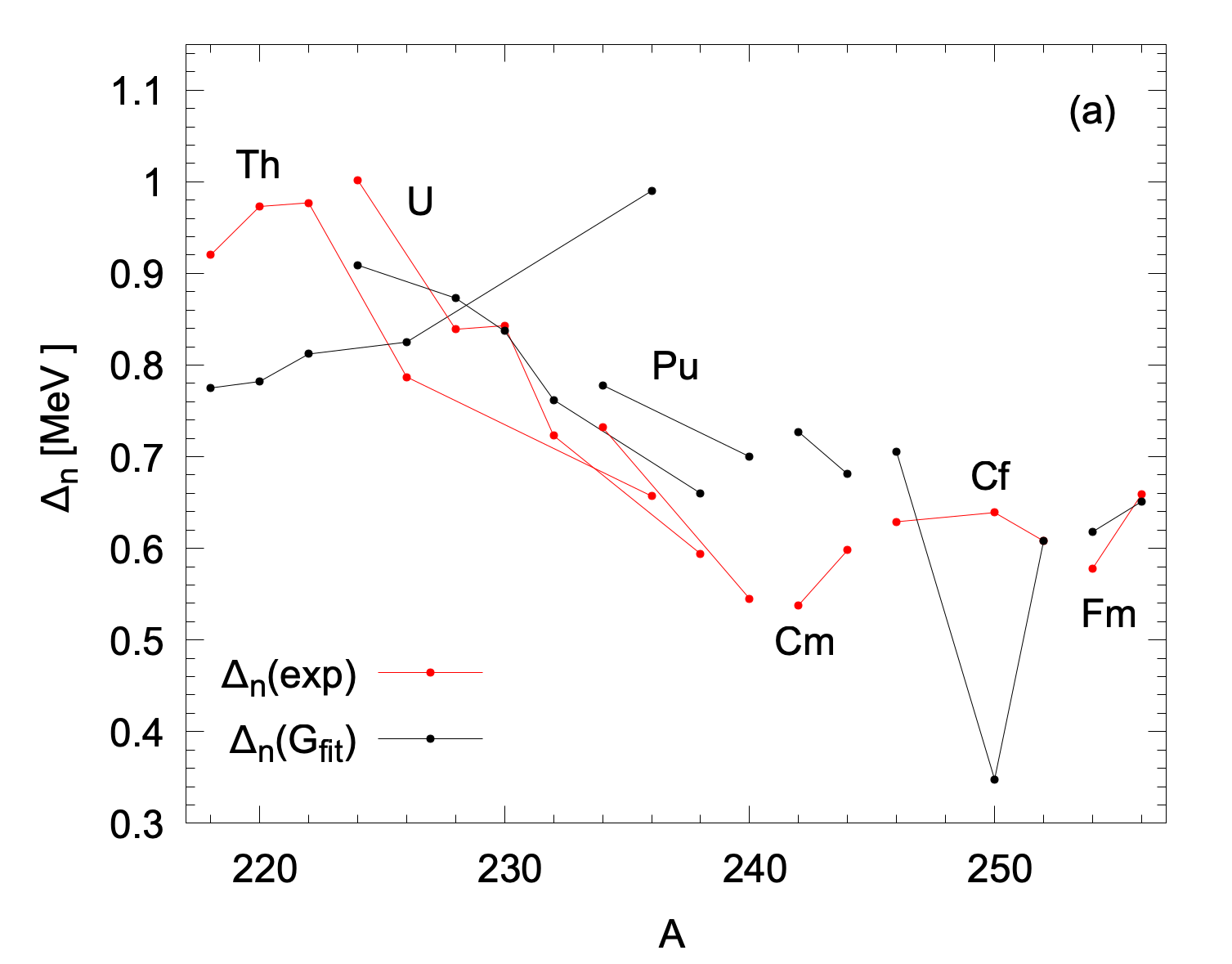}
      \hskip-0.5cm
      \includegraphics[scale=0.172]{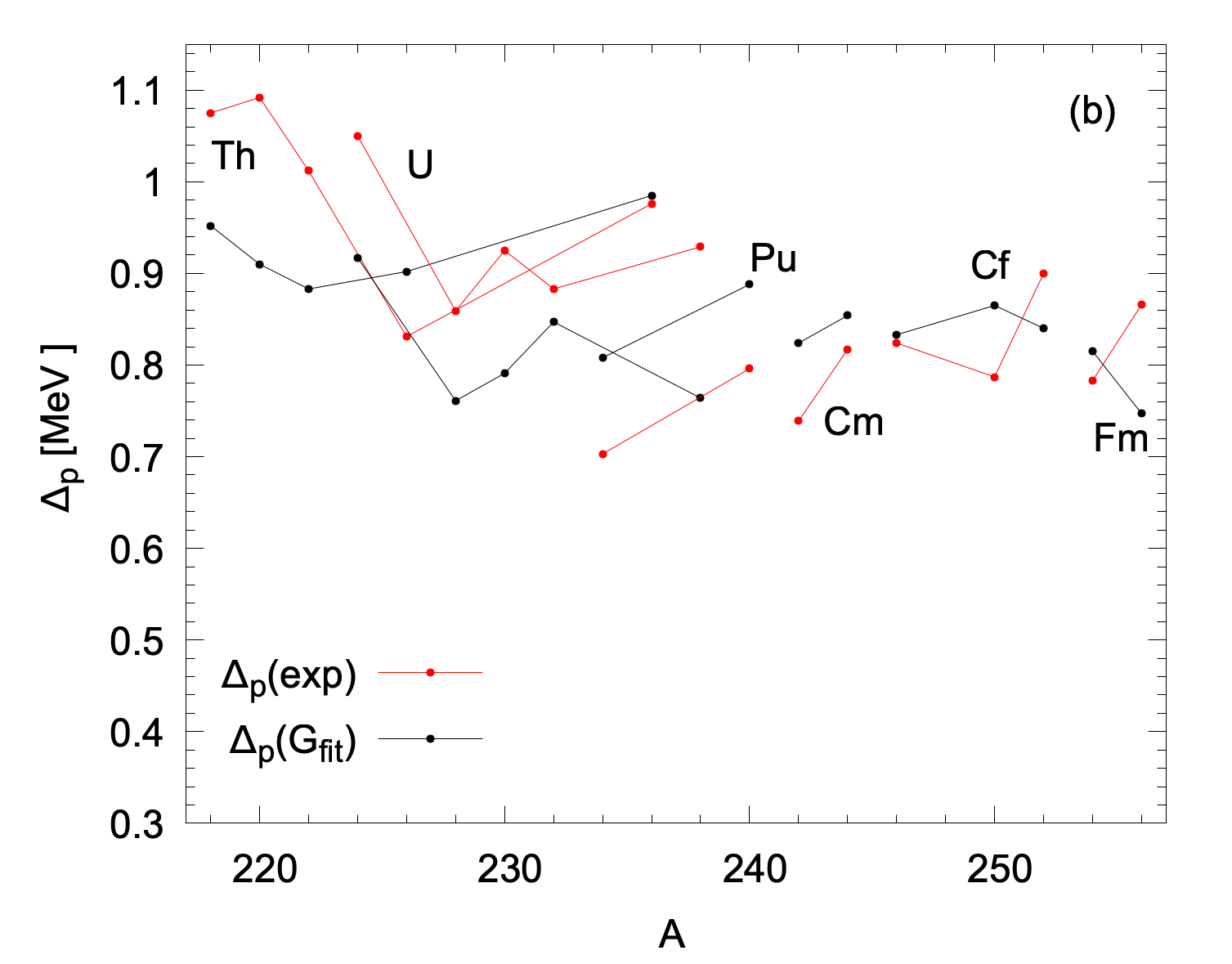} 
      \vskip 0.5cm  \hskip 0.5cm
      \includegraphics[scale=0.172]{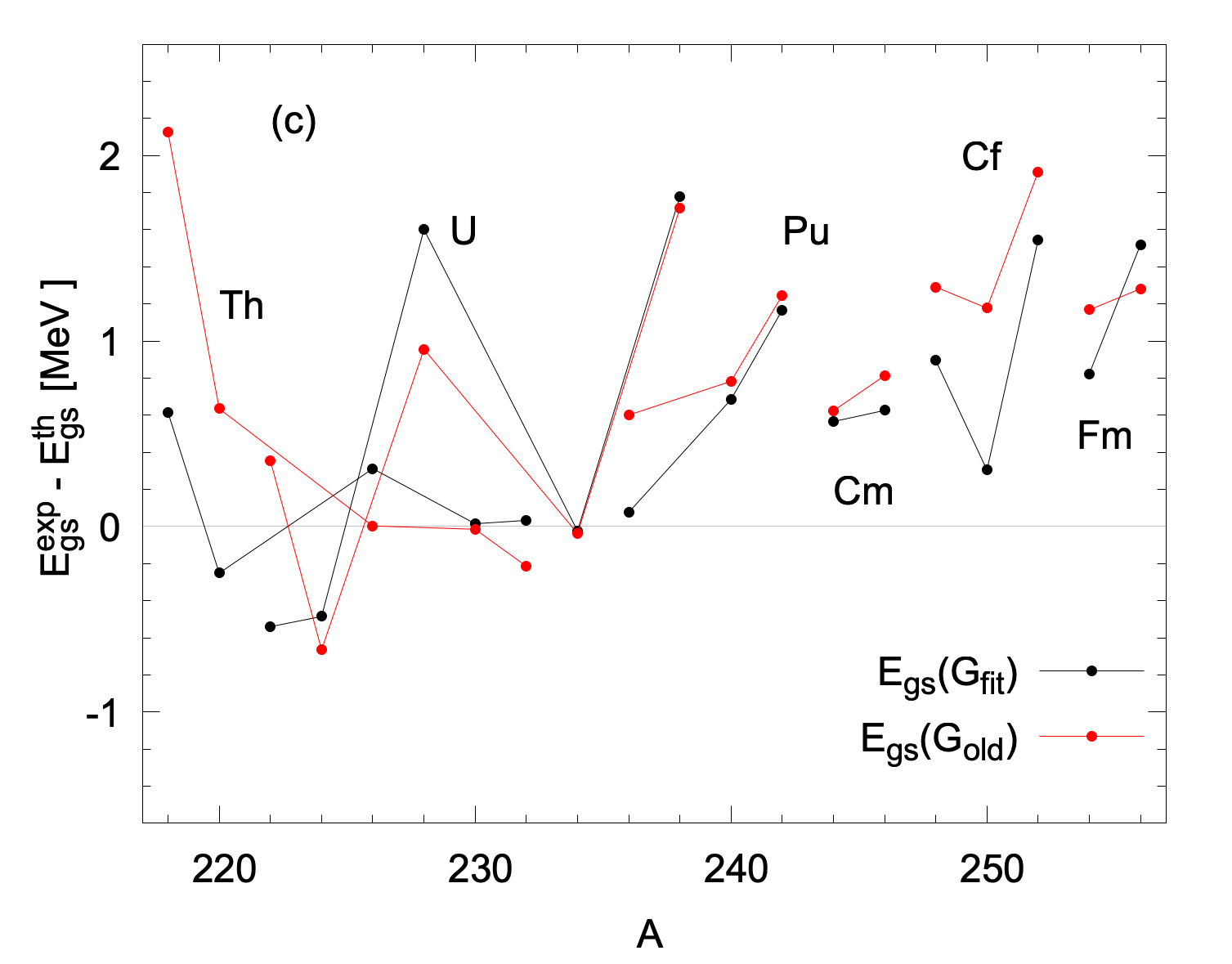}
    \caption{Comparison between calculated (black triangles) and empirical (red dots) neutron (a) and proton (b) pairing gaps for the discussed isotopic chains from $Z=90$ up to $Z=100$. For a better visibility we present for a given value of the mass number $A$ only the one isotope, for which the discrepancy between the theoretical and experimental values is the largest. Panel (c) displays the largest differences between experimental and calculated ground-state masses, evaluated with the pairing strength of Eq.\ (\ref{Gstas}) (red dots) and the one obtained from Eq.~(\ref{Gqfit}) (black triangles).}
\label{fig.06}
\end{figure*}

The energy gap $\Delta_q$, ($q\!=\!n$ or $p$) for neutrons or protons produced by the pairing interaction can be expressed as $\Delta_q\!=\!E^{(q)}_{int}/2$, with $E^{(q)}_{int}$ the interaction energy between two nucleons of type $q$. 
For a given nucleus with particle numbers $N_q = N$ or $Z$, and the corresponding separation energies $S(N_q)$ 
\begin{equation}
  E^{(q)}_{int}=S(N_q)-\frac{1}{2}[S(N_q+1)+S(N_q-1)],
\end{equation}

the pairing gaps are easily expressed in terms of the empirical mass excesses $B(N_q)$ taken from Ref.~\cite{mass_tables} in the following way: 
\begin{equation}
  \Delta_q^{(exp)} = \frac{1}{4} \left[2B(N_q) - B(N_q+1) - B(N_q-1) \frac{}{}\!\right].
\label{Del_exp}\end{equation}

The pairing strengths $G$ can therefore be deduced, for the here considered 39 heavy nuclei with $Z=90-100$ for which the ground-state masses are known, by requiring that the pairing gaps $\Delta_q(G)$ obtained in the BCS approach (including the particle-number projection) are found as close as possible to their empirical values calculated through Eq.~(\ref{Del_exp}).
One then requires the following expression 
\begin{equation}
  \sum_{\rm set} |\Delta_q^{(exp)} - \Delta_q(G)| 
\label{miniG}\end{equation}
to be minimal, where the sum runs over the set of the considered nuclei.
\\[ -2.0ex]

In order to facilitate the above discussed calculation, one usually tries to find, in practice, a simple 
analytical expression, depending on $N,Z$, which is able to reproduce the pairing strength $G$ for both protons and neutrons in the best possible way. Among many such phenomenological expressions, one which has proven quite successful may be written in the following form:
\begin{equation}
  G\,A = g_0 + g_1\,(N-Z) \,.
\label{Gqfit}\end{equation}
This expression depends on only two free parameters $g_0$ and $g_1$ which are fitted to the value of $G$ that renders the expression of Eq.~(\ref{miniG}) minimal. 

The optimal values for these two parameters have been found for our sample of 39 actinide nuclei 
to be $g_0 = 18.35$ MeV and $g_1 = 0.103$ MeV for protons and $g_0 = 24.1$ MeV and $g_1 = -0.135$ MeV for neutrons.
\\[ -2.0ex]


  


The quality of this fit is visualized in Fig.~\ref{fig.06} (a) and (b) where the values of pairing gaps calculated using Eq.~(\ref{Gqfit}) are compared  with the empirical ones obtained from Eq.~(\ref{Del_exp}). 
To make this comparison more transparent, we chose to present for each isobaric chain only the one isotope, for which the discrepancy between the theoretical and experimental values is the largest.  
One finds that the largest deviation for neutrons does not exceed 0.35 MeV ($^{236}$Th and $^{250}$Cf) and for protons is always lower than 0.2 MeV. It is worth to point out that on average, the largest deviations from the empirical pairing gaps for both types of nucleons reach around $0.12$ MeV. 
Panel (c) of the Fig.~\ref{fig.06} presents, for the nuclei of panels (a) and (b),
the macroscopic-microscopic ground state energy, relative to the experimental data, with the pairing corrections obtained using the prescription of  Eq.~(\ref{Gqfit}) (black triangles)  and a previous fit of the pairing strength (red dots) within the same projected BCS-like formalism as presented above (see \cite{GfitS} and references therein),  where the nucleon number dependence of $G$ is given by: 
\begin{equation}
G_{q}\cdot N_{q}^{2/3}=g^{(0)}_{q},\quad q=\{n,p\}.
\label{Gstas}
\end{equation}

The only parameter $g^{(0)}_q$ in this 
parametrisation of the pairing strength is chosen as $g^{(0)}_{q} \!=\! 0.28\hbar\omega_0$ with a value of $\hbar \omega_0=41/A^{1/3}\,$MeV of e.g. Ref.~\cite{Gqfit41}, 
widely used in macroscopic-microscopic calculations, and common for both protons and neutrons. 

As demonstrated in Fig.~\ref{fig.06}, our new pairing-strength fit, Eq.~(\ref{Gqfit}), gives, for most of the nuclei in question, an overall better reproduction of experimental ground-state masses as compared to its older version (\ref{Gstas}), with the exception of the uranium isotopes
$^{222,226,228}$U and of $^{256}$Fm isotopes. Only in $^{228}$U does the discrepancy between the two models reach half an MeV to the disadvantage of the new fit.

Taking into account the above considerations, one can now write down the total energy of the nuclear system in the macroscopic-microscopic approach simply as
\begin{equation}
  E(N,Z,def) = E_{LSD} + \sum_{q} \left[\delta E^{(q)}_{shell} + \delta E^{(q)}_{pair} \right]
\label{dE}\end{equation}
with the shell and pairing corrections $\delta E^{(q)}_{shell}$ and $\delta E^{(q)}_{pair}$ $(q=n,p)$ being given by Eqs.\ (\ref{ESH}) and 
(\ref{Epaircor}), respectively.
\\[ -2.0ex]

Using the above prescription, we determine the nuclear energy as function of the deformation parameters $\eta$, $q_2$, $q_3$, $q_4$  introduced in Section II.A which stand respectively for the non-axiality, elongation, left-right asymmetry and neck formation of the nuclear shape. 
The collection of all these energy points constitutes what we call the deformation energy or potential energy surface (PES) of a given nucleus on a discrete four-dimensional mesh.
We have chosen a step length of $\Delta q_2=0.05$ for the elongation parameter $q_2$ and a step length of $\Delta q_j=0.03$ for the other 3 deformation parameters with a total mesh size of 
$n_2 \times n_3 \times n_4 \times n_\eta = 60 \times 8 \times 15 \times 8=57600$ nodes. 
We have verified that within such a mesh size we are able to describe with a good enough accuracy all physically relevant effects, like local minima, saddle points, the formation of valleys and ridges.

\section{Multidimensional WKB method}

The WKB method is a semi-classical approximation which is widely used in quantum mechanical problems to find an approximate solution of the Schr\"odinger equation implying a potential barrier that a particle has to overcome. The main assumption is that, 
under the influence of the potential, the particle wave function can still be expressed in terms of a plane wave, but with a momentum $k(x)$ which is position-dependent and slowly varying with $x$.

\subsection{Lifetimes for spontaneous fission} 
In our approach we have used 
a multidimensional version of the above characterized WKB approximation to calculate the lifetime of a nucleus undergoing spontaneous fission. This approach has been widely used in nuclear physics for fission and particle or cluster emission to determine the penetrability of a potential-energy barrier defined in a multidimensional deformation space. In the following, the standard one-dimensional WKB method will be generalised to the case of a four-dimensional deformation space, where the deformation variables are the Fourier parameters $q_i$ introduced in Eq.~(\ref{a_to_q}).

The first step to obtain a good-quality estimate of the lifetime of a system undergoing spontaneous fission is to search for the so-called ``{\it least-action path}'' (LAP) leading to fission in our 4-dimensional PES that a nucleus would have to follow on its way to a splitting into fission fragments. Such an approach treats a fission event as a dynamical process, characterized by the collective motion of a large number of nucleons tending to elongate the nuclear shape starting from some initial state, like the nuclear ground state until the scission configuration is reached. Please note that the collective space in which the fission process is simulated can generally be multidimensional, curvilinear and non-Euclidean. 
 In the framework of our present approach the dynamical path to fission actually proceeds in the four-dimensional space defined by the 
($q_1\equiv \eta,q_{2},q_{3},q_{4}$) Fourier deformation parameters defined through Eqs.~(\ref{eta})-(\ref{transf}).
We have investigated that when 
triaxiality
$q_1$ is taken into account, one can observe along the least-energy path 
a slight (up to 1 MeV) lowering of, in particular, the inner fission barriers height. 
\\[ -2.0ex]

One should keep in mind that in the here presented approach the energy $E(q_1,q_2,q_3,q_4)$ is obtained in the macroscopic-microscopic model, where the shell corrections in (\ref{dE}) are determined in the Strutinsky method, and the correction for the residual pairing interaction in the BCS approximation with projection onto good particle number obtained in the GCM approach (see Ref.\ \cite{GCM+GOA}). In both these methods, single-particle states of a folded-Yukawa mean-field potential \cite{ldens} are used.

\subsection{Least-action fission path}
The action in the above introduced 4-dimensional deformation space $\{q_1,q_2,q_3,q_4\}$ can be represented through the following integral:
\begin{equation}
\begin{split}
S = \!\!\!\!\!\! \int\limits_{q_{2}^{(g.s.)}}^{q_{2}^{(exit)}} \!\!\!\!\! dq_2\sqrt{\frac{2}{\hbar^2}\sum_{ij=1}^4\beta B_{ij}(\{q_k\})
    \big[{\mathcal E}\!-\!E(g.s.)\big]\frac{\partial q_i}{\partial q_2}\,\frac{\partial q_j}{\partial q_2}},
\end{split}
\label{LAP}
\end{equation}
where ${\mathcal E}\!=\!E(\{q_k\})$ and $E(g.s.)$ stand respectively for the potential energy of any configuration along the fission path and at the ground state deformation. The integration extends from the nuclear ground state deformation up to a so-called ``$\!${\it exit point}$\,$'' which has the same energy as the ground state ($E_{exit}=E_{g.s.}$). 
In the 4-dimensional deformation space one can obviously find many such turning points which fulfill this condition. The problem now consists in the identification of the one particular exit point and the corresponding path leading to it, which renders the action integral minimal.
At first sight, one may have the impression that the action integral in Eq.\ (\ref{LAP}) is computed only along a one-dimensional fission path L($q_2$) instead of being defined in the multi-dimensional collective space. That is, however, not the case, since that search for the minimal action path is carried out in the full 4-dimensional deformation space. A similar kind of approach has also been used e.g.\ in Refs. \cite{LS99,LBP03} to calculate the LA integral in a multi-dimensional deformation space.

\begin{figure*}[th!]
    \includegraphics [scale=0.15] {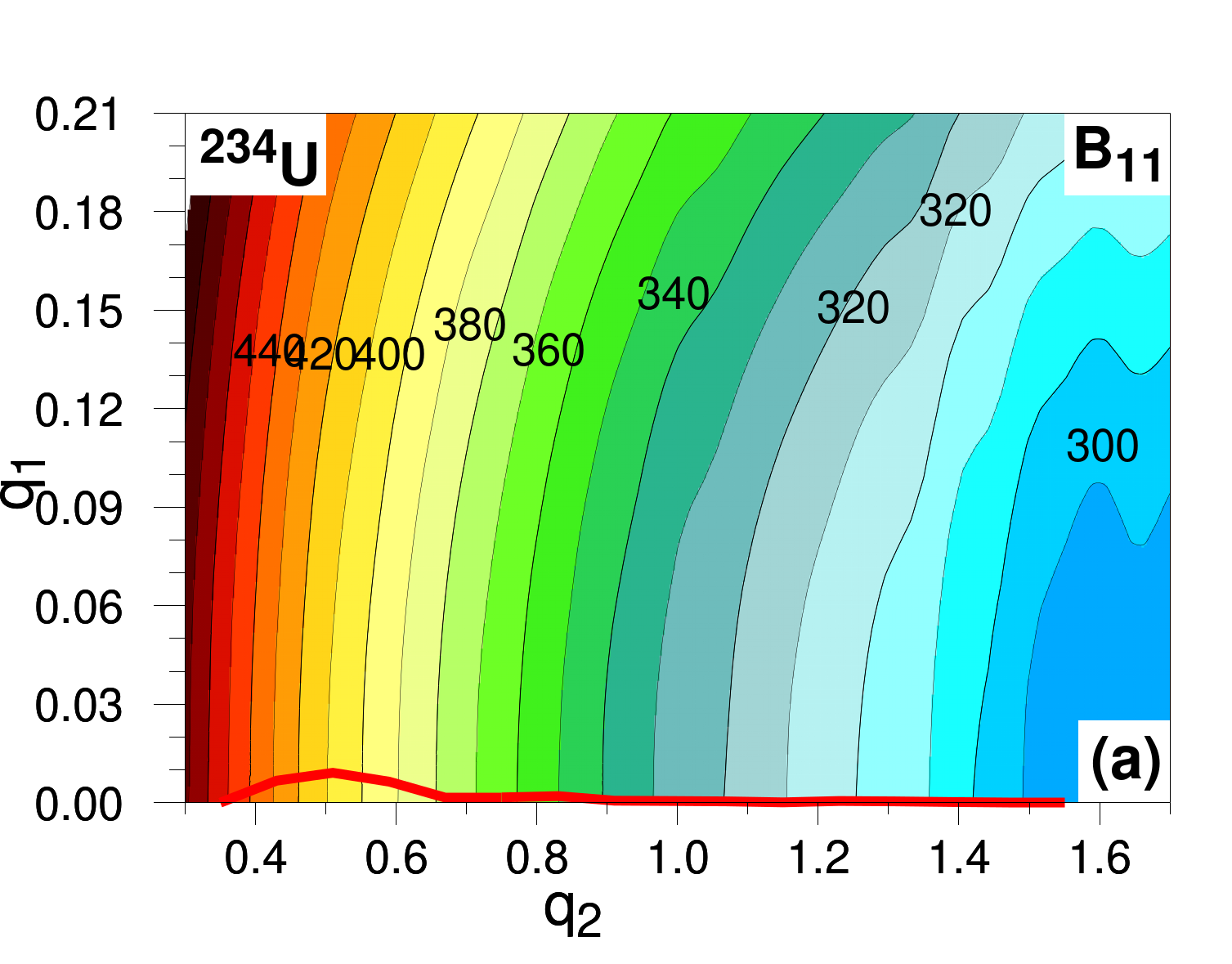}
    \includegraphics [scale=0.15] {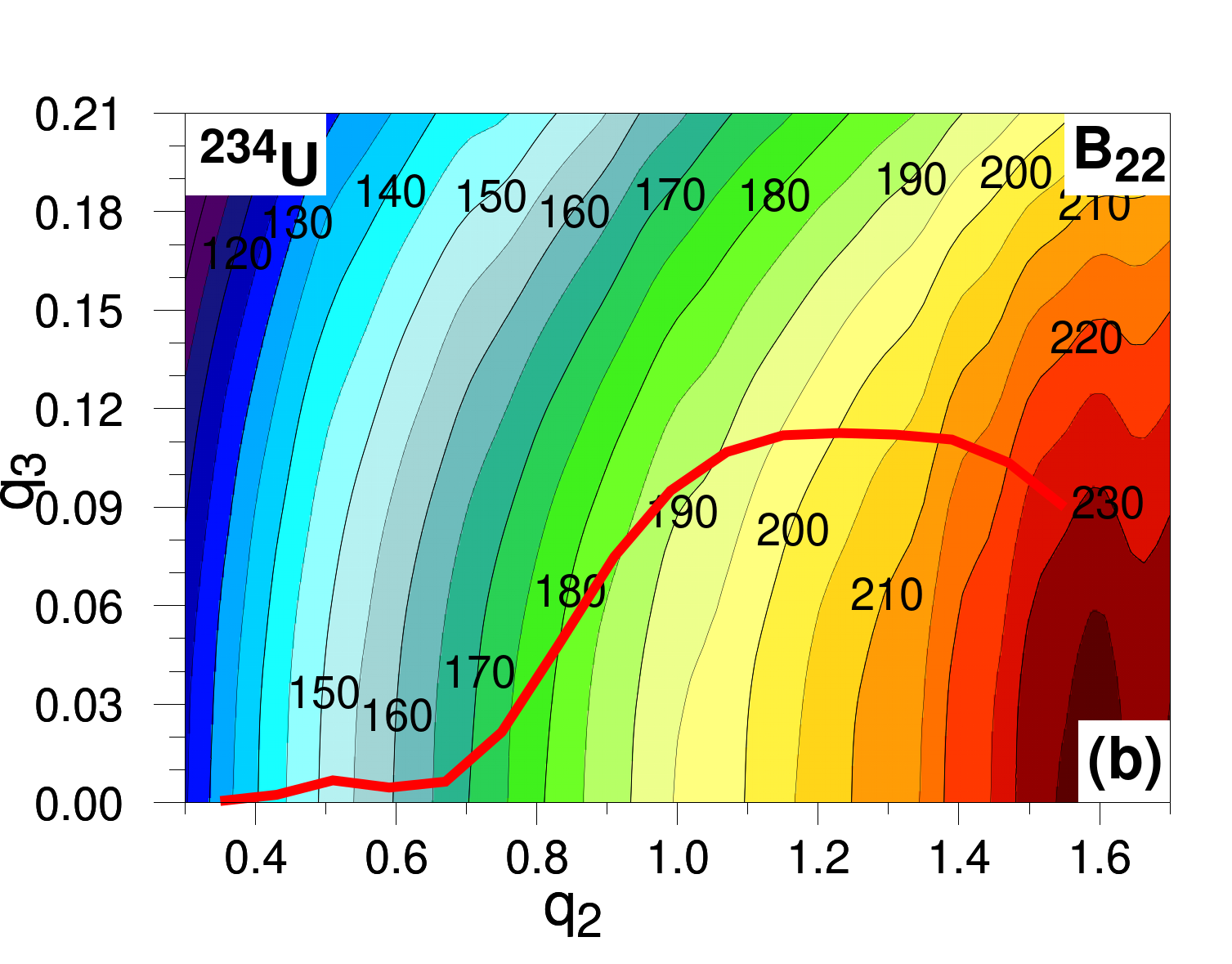}\\
    \includegraphics [scale=0.15] {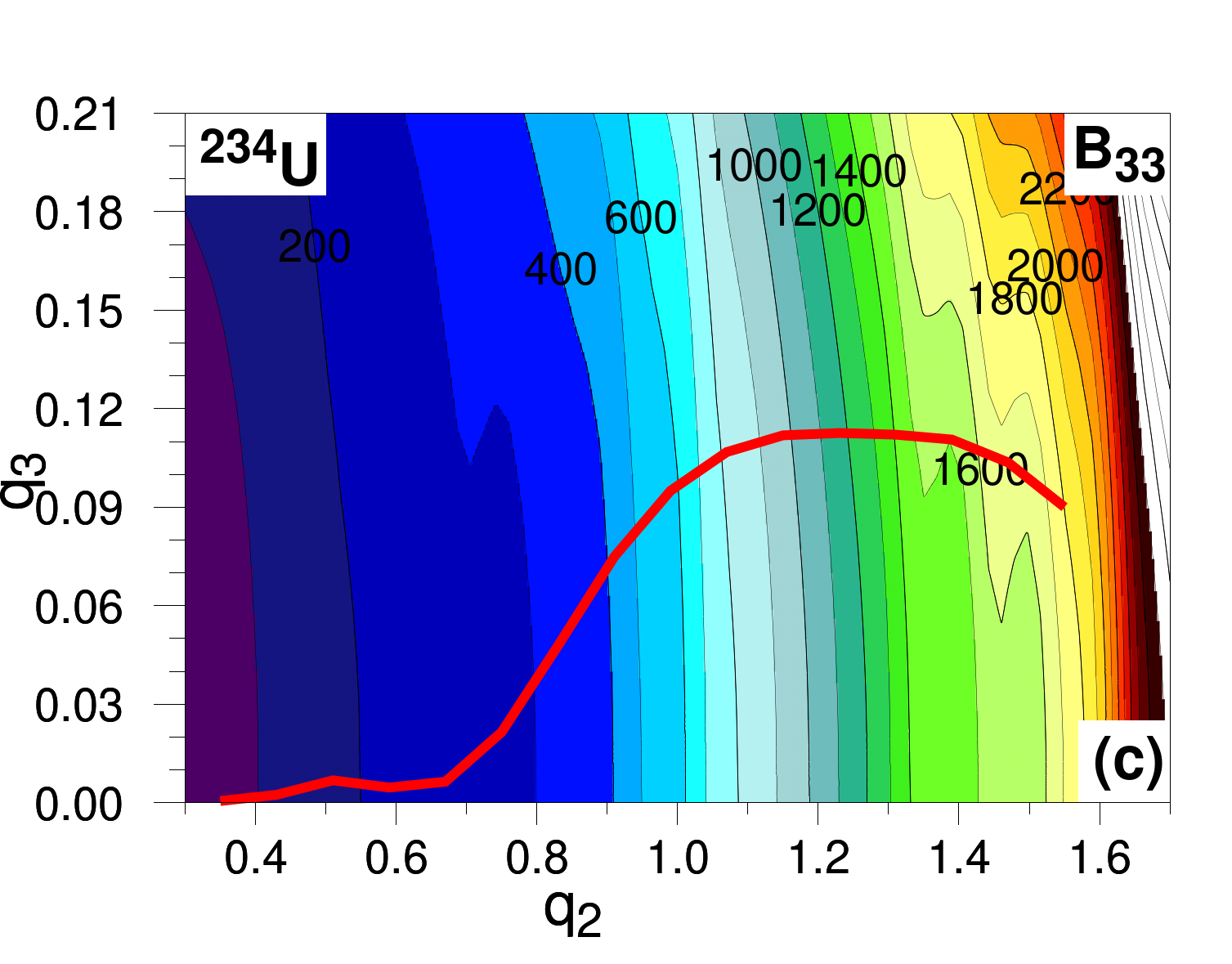}
    \includegraphics [scale=0.15] {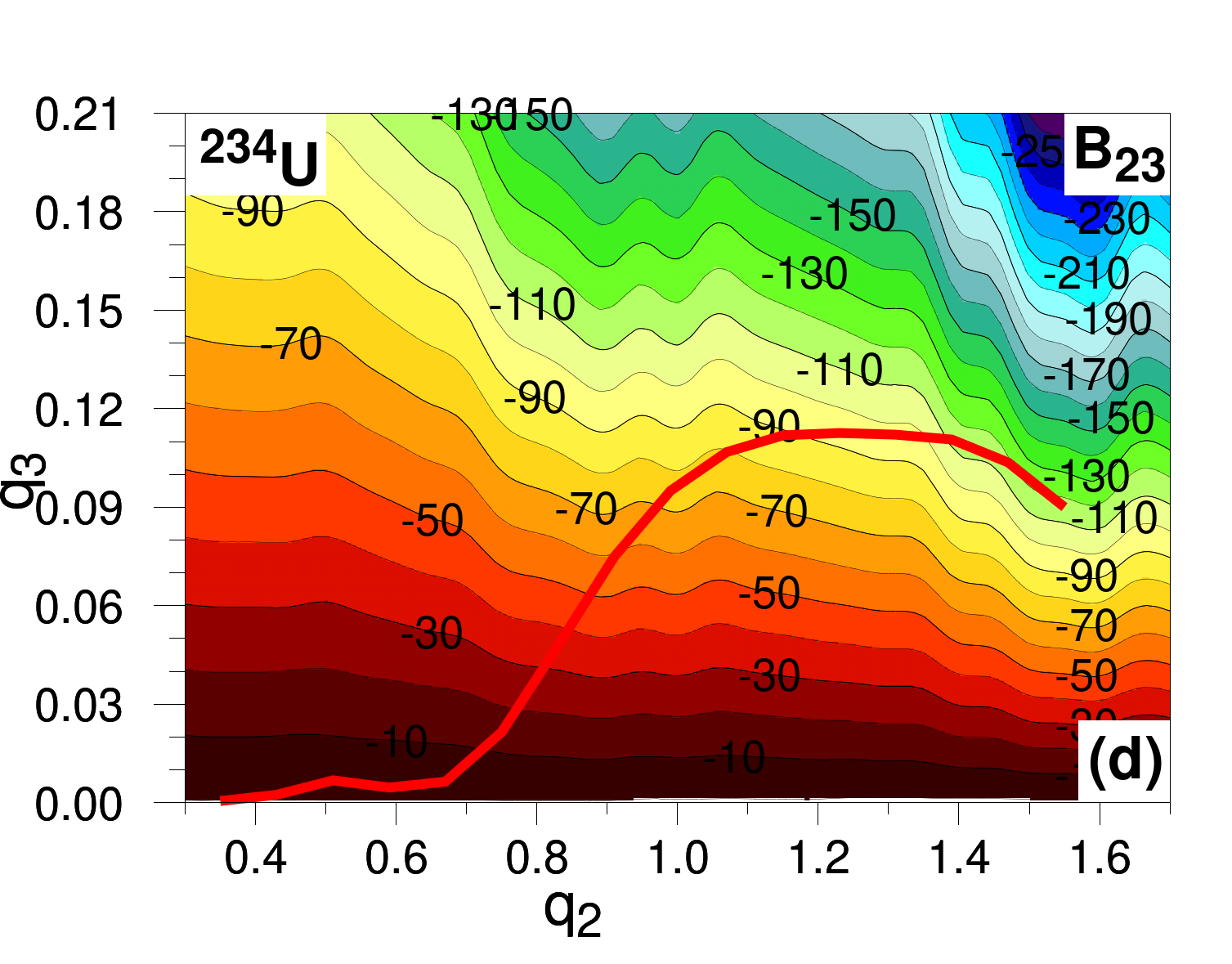}  

    \caption{Hydrodynamical mass tensor components in the $(q_2,q_1)$ (a) and $(q_2,q_3)$ (b-d) deformation planes.
             The LAP is indicated by the solid red line.}
\label{Bij_maps}\end{figure*}

The deformation-dependent quantity $B_{ij}(\{q_k\})$, with indices $i,j$ ($i=1,2,3,4$) referring to the pair of shape parameters $(q_i,q_j)$, is the irrotational flow inertia tensor in the Werner-Wheeler approximation, of Refs.~\cite{DSN76, BN19} and presented in Fig.~\ref{Bij_maps} for the diagonal components $B_{11},\, B_{22}$ and $B_{33}$ as well as for the off-diagonal component $B_{23}$ projected onto the $(q_2,q_1)$ and $(q_2,q_3)$ plane respectively. 
The values of the remaining two deformation parameters $q_3,\,q_4$, respectively $q_1,\,q_4$
are adjusted in such a way that they minimize the action integral (\ref{LAP}) along the least-action path.

Please notice that the deformation dependent parts of this purely macroscopic hydrodynamical mass tensor is essentially identical for all nuclei and has simply to be multiplied by a scaling factor proportional to $A^{5/3}$, (see, e.g.\  Ref.~\cite{BN19}) to obtain the proper value for a given particular nucleus.

As can be seen from Fig.~\ref{Bij_maps}, the components $B_{22}$ gradually increases with the elongation coordinate $q_2$, but is only weakly dependent on the mass-asymmetry parameter $q_3$ in the region where it would have the largest influence on the action integral (\ref{LAP}), namely the barrier region around $q_2 \leq 1$. The other crucial inertia component, $B_{33}$, changes relatively slowly with $q_2$ in this region (below $q_2 \!\lesssim\! 1$), but increases dramatically in the vicinity of the scission configuration ($q_2 \gtrsim 1.5$). 
In contrast, the component $B_{11}$ decreases with elongation $q_2$, but stays almost constant when $q_1$ increases. Since the WKB action integral is determined by an interplay between potential energy and inertia tensor, we may thus conclude that it is the potential energy gradient in $q_1$ direction which will mostly contribute to the final value of the action integral and the impact of $B_{11}$ is small in the presented case. 
Please notice also that the absolute values of $B_{33}$ are much larger than the ones of $B_{22}$, and thus may contribute substantially to the total action (\ref{LAP}) whenever a derivative $\frac{dq_3}{dq_2} \neq 0$, comes into play. This essentially happens if the fission path leads to a mass-asymmetric division, even if $q_3$ stays about constant in its final stage. If, in turn, the exit point is mass symmetric ($q_3=0$), the observed local variations of $q_3$ along the LAP are too weak to significantly contribute to the action integral. Please notice also that for the here studied nuclei the fission path usually starts heading towards mass asymmetric deformations around the second minimum ($q_2 \approx 0.7-0.8$) as this is shown for the cases of $\,^{234}$U or $\,^{252}$No in Fig.~\ref{LAP_plots}. It is interesting to note that changes of the off-diagonal mass component $B_{23}$ shown in Fig.~\ref{Bij_maps}(d) favour asymmetric fragmentation.\\[-2.0ex]

The parameter $\beta$ in front of the $B_{ij}$ mass tensor in (\ref{LAP}) gives the possibility of re-scaling its six independent components all together in order to reproduce within a couple of orders of magnitude the measured actinide half-lives. Obviously, such an operation does not touch the relative values between the tensor components what, in addition to the reliability of the PES, is relevant for a realistic determination of the course of the LAP, and thus of the resulting action value.  Let us also recall that the differences in 
the 
inertia components evaluated within several available macroscopic or microscopic approaches may even differ by 
as much as 
one order of magnitude \cite{robmass}.  

One of the distinguishing features of the macroscopic hydrodynamical mass tensor used here as compared to its microscopic 
(e.g. cranking model)  
counterpart lies in the fact that the latter is often a rapidly fluctuating function of deformation, caused mainly by the microscopic shell effects.
These local variations are, to some extent, smoothed out by the method of the least-action trajectory itself, where the corresponding fission path tends to omit states associated with a sudden 
change of the potential energy or the inertia. This has a clear impact on the stability of the numerical search for the minimum of the action integral (\ref{LAP}) in a multi-dimensional space of variational parameters. 
For a comparison, we have applied, in addition, another efficient prescription of the collective inertia effects simulated by the so-called phenomenological mass parameter $B(R_{12})$, expressed in units of the reduced mass $A_L A_R/(A_L+A_R)$, with $A_L$ and $A_R$ being respectively the mass number of the left and right nascent fission fragments (see, e.g. Ref.~\cite{DNS76}). 
\begin{equation}
  B(R_{12})=1+k\frac{17}{15}\exp\bigg[\lambda \, (R_{12}^{({\rm sph})} - R_{12})\bigg].
\label{masa}\end{equation}
The above phenomenological mass depends on a single parameter $R_{12}\!=\!R_{12}(q_2,q_3,q_4)$ 
(in units of the radius $R_0$ of the spherical shape) 
describing the evolution towards fission and which is given by the centers-of-mass distance 
(for a spherical shape one has $R_{12}^{({\rm sph})} \!=\! 0.75\,R_0$)
of the nascent fission fragments. The parameter $\lambda \!=\! 0.408/R_0$ describes the {\it descent rate} of the exponential function.
For this purpose, in contrast to the calculations with full hydrodynamical mass tensor, the 3D total potential energy function $E(q_2,q_3,q_4) = E(q_1^{0},q_2,q_3,q_4)$, Eq.\ (\ref{dE}) is used, 
where $q_1^{0}$ is the nonaxiality deformation parameter which minimizes the full 4D potential energy $E(q_1,q_2,q_3,q_4)$ at at given point in the 3D $(q_2,q_3,q_4)$ space. 
%
%

%
%
\begin{figure*}[ht!]
      \includegraphics[scale=0.15]{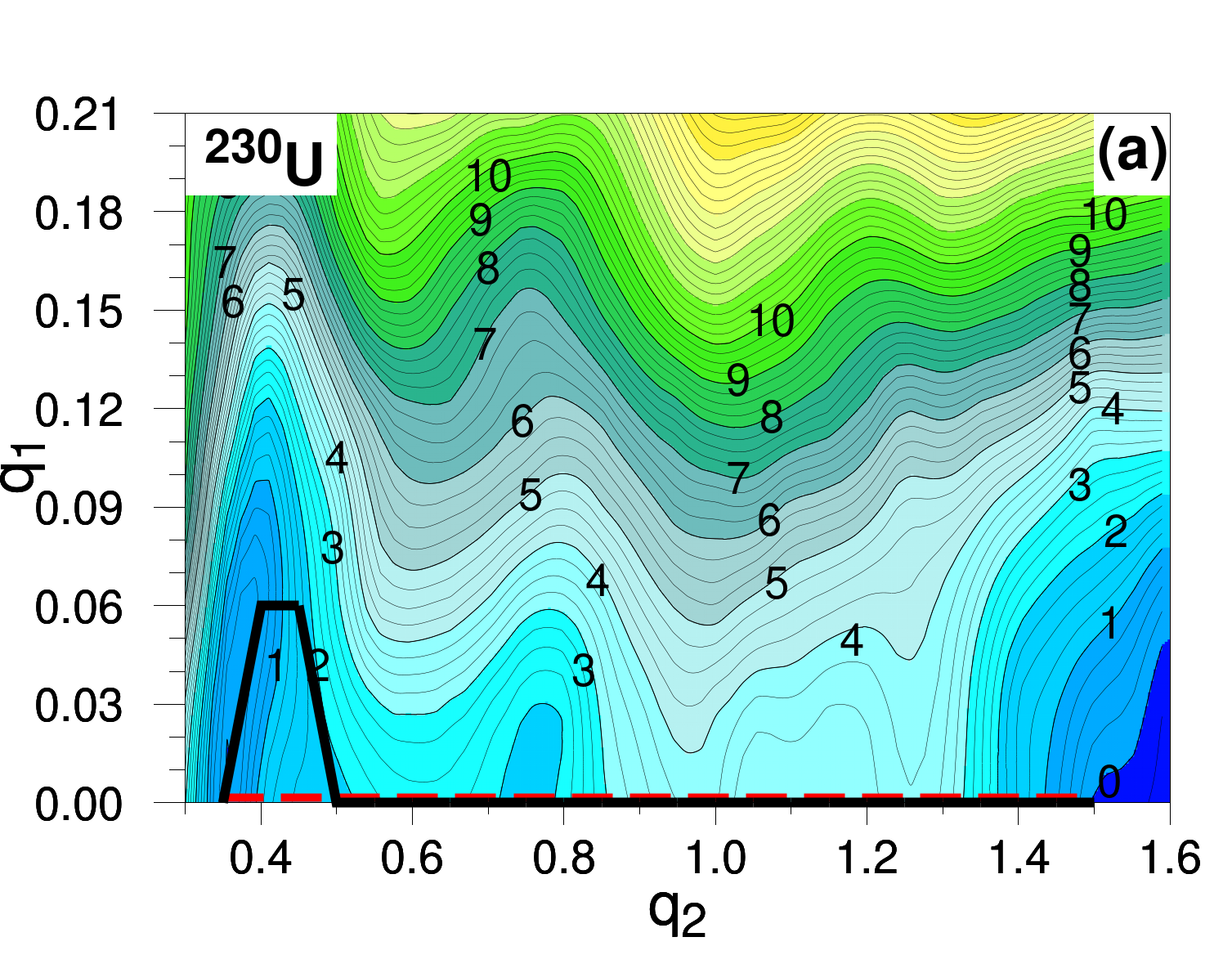}
      \includegraphics[scale=0.15]{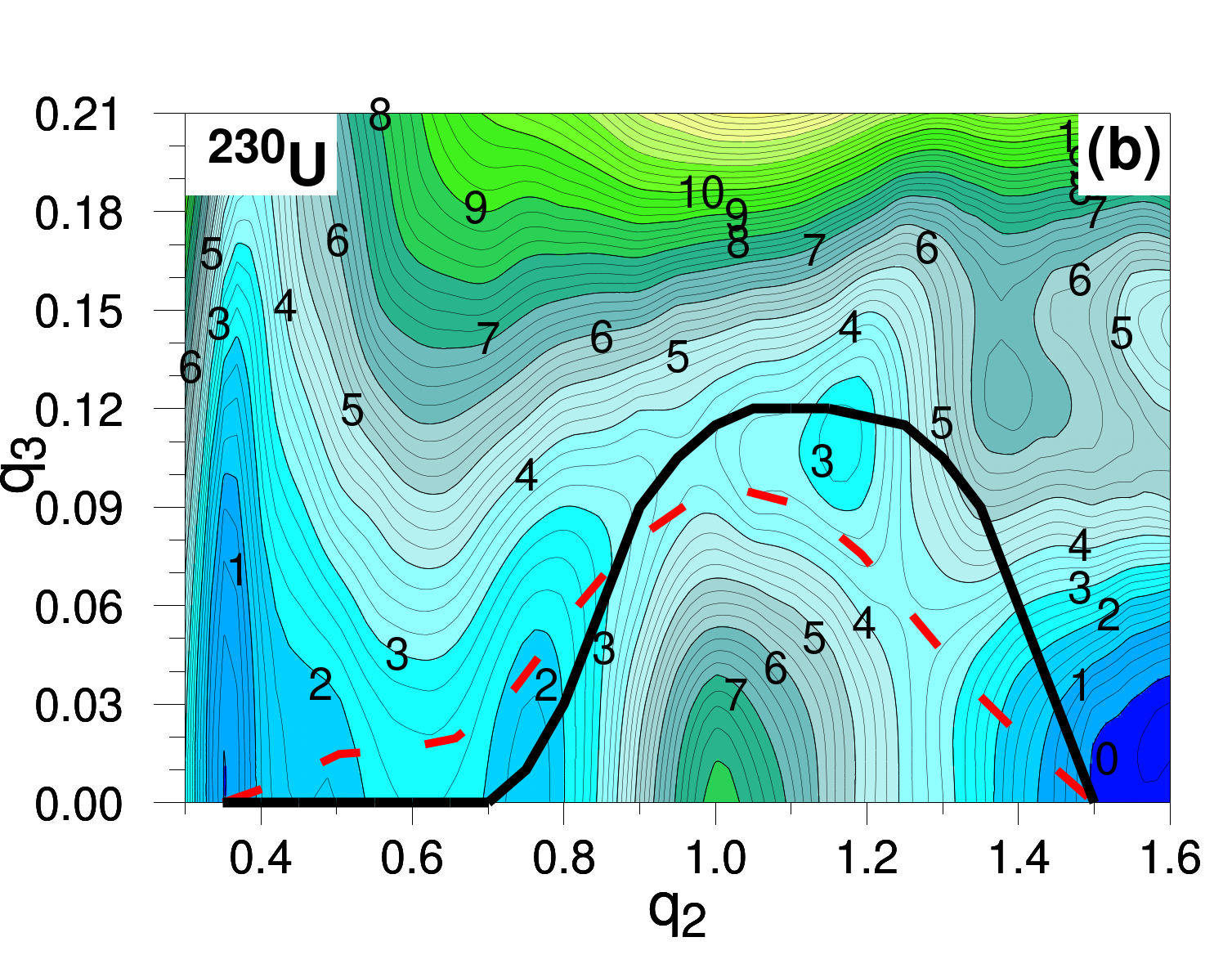}\\
      \includegraphics[scale=0.15]{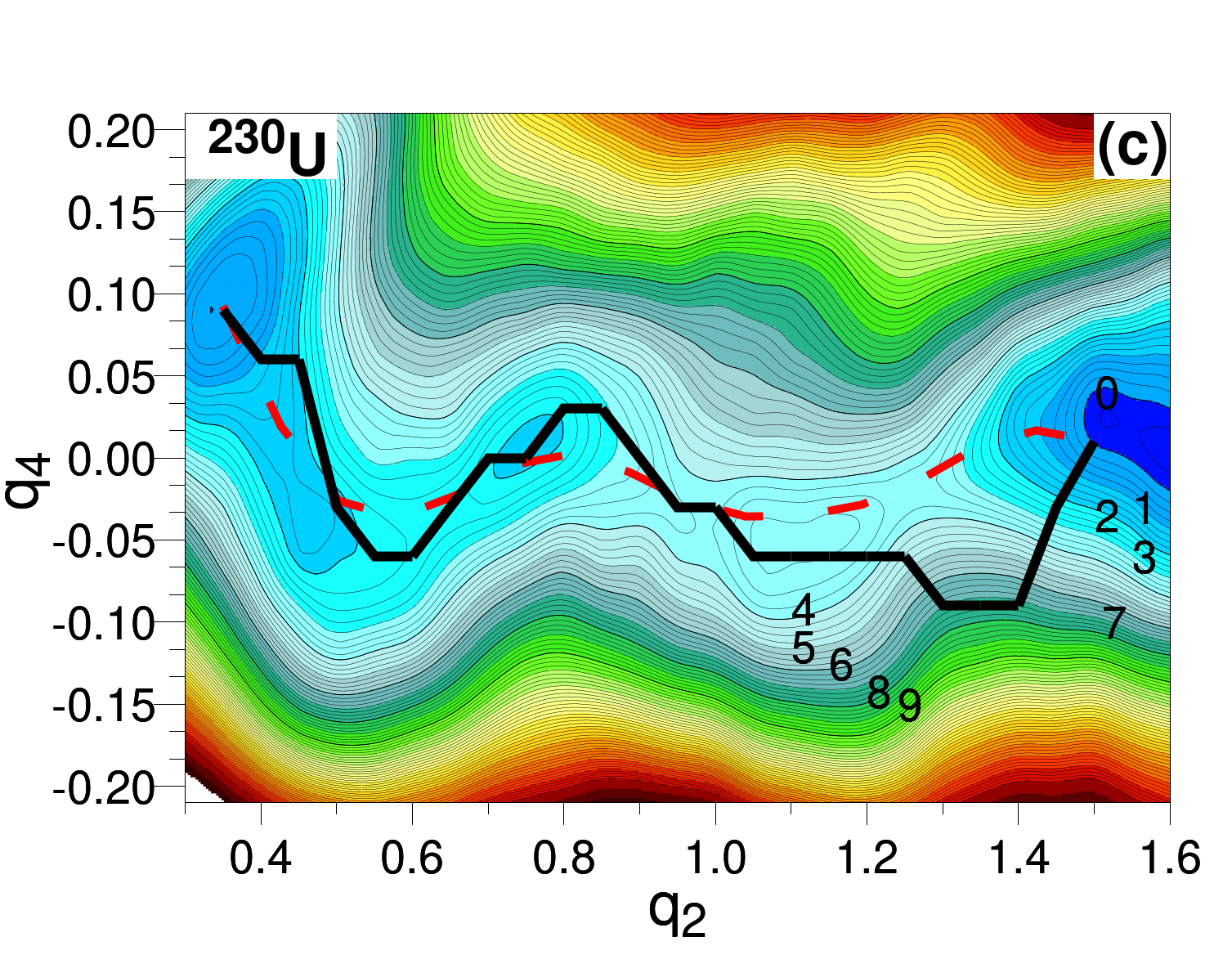} 
      \includegraphics[scale=0.15]{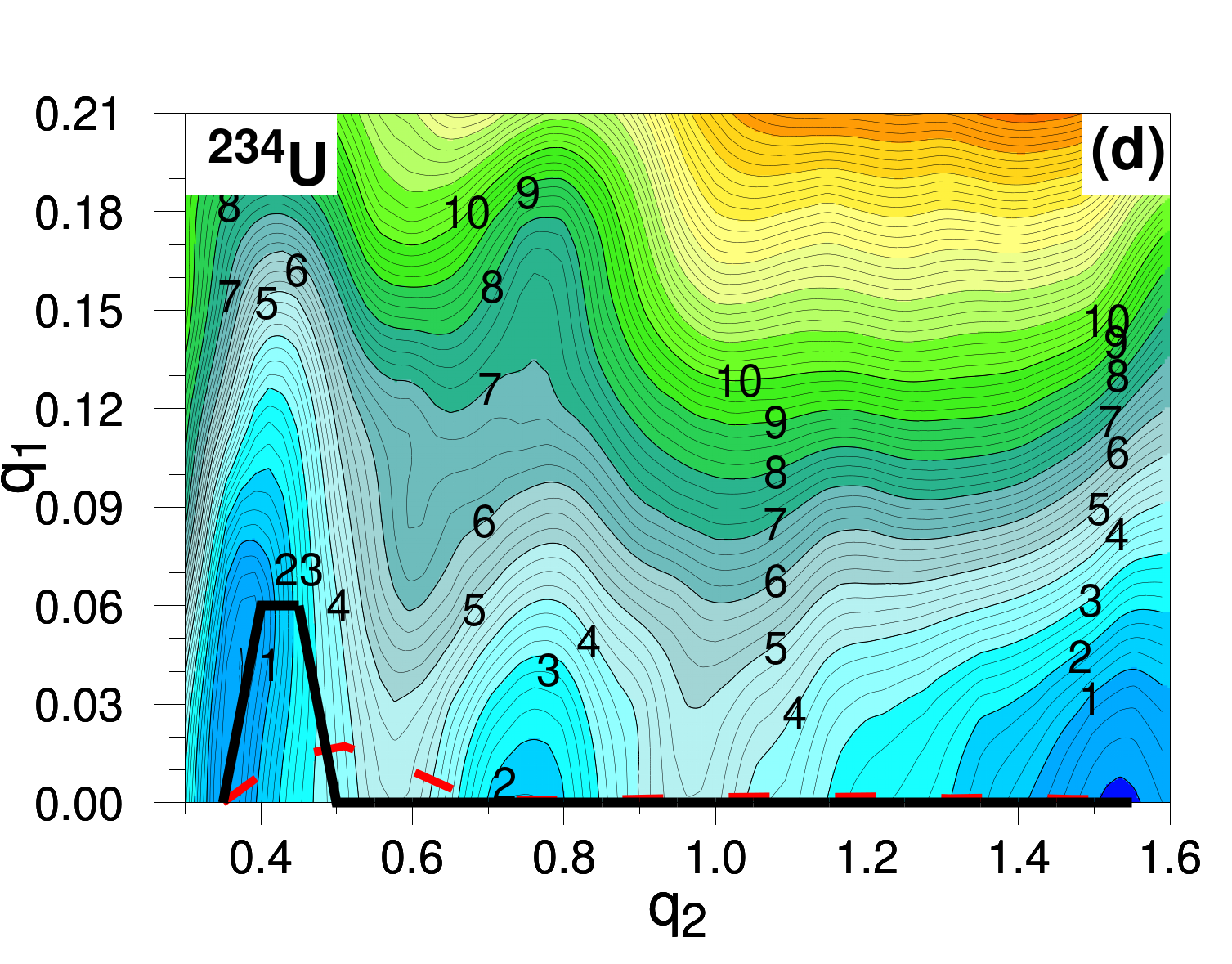}\\
      \includegraphics[scale=0.15]{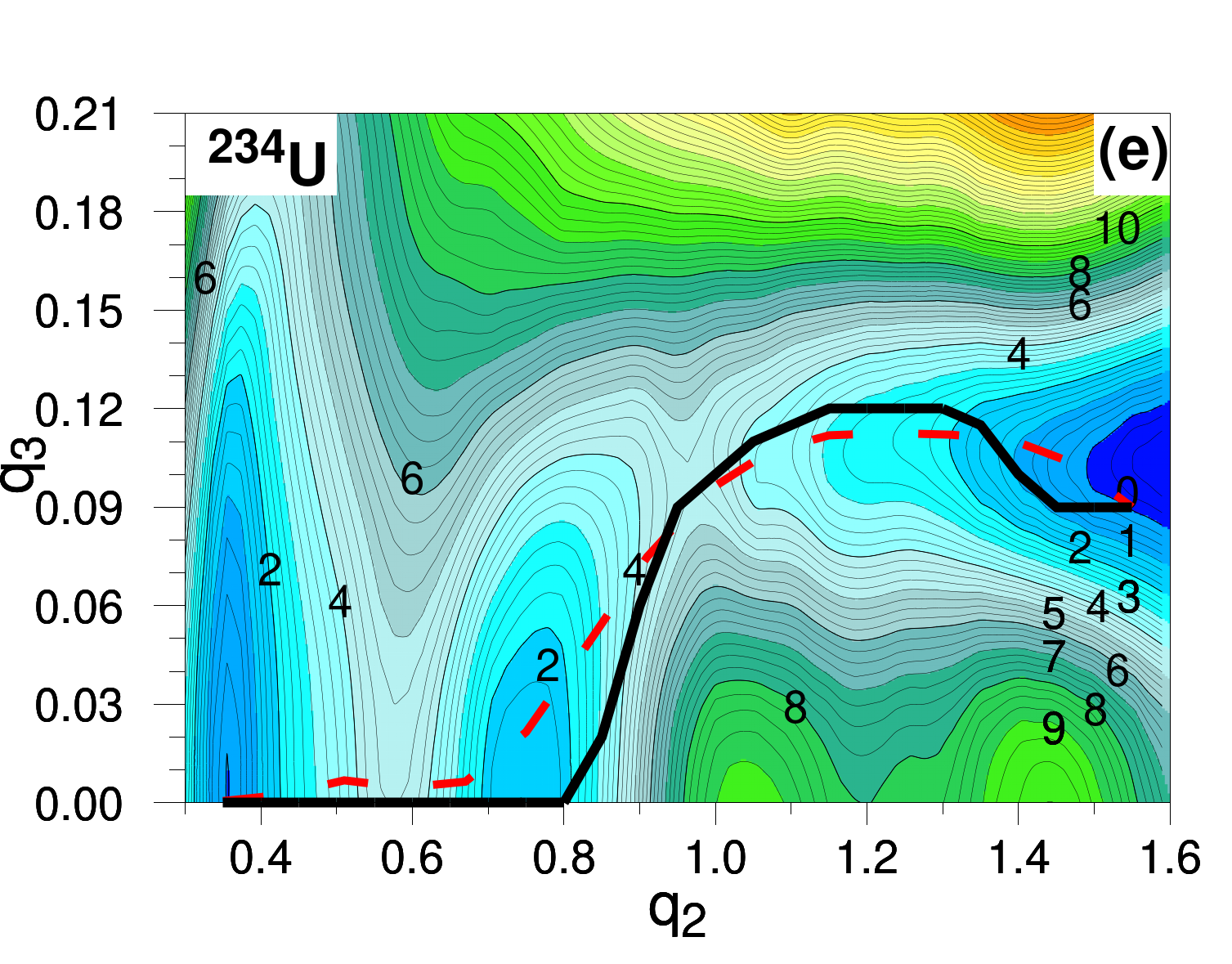}     
      \includegraphics[scale=0.15]{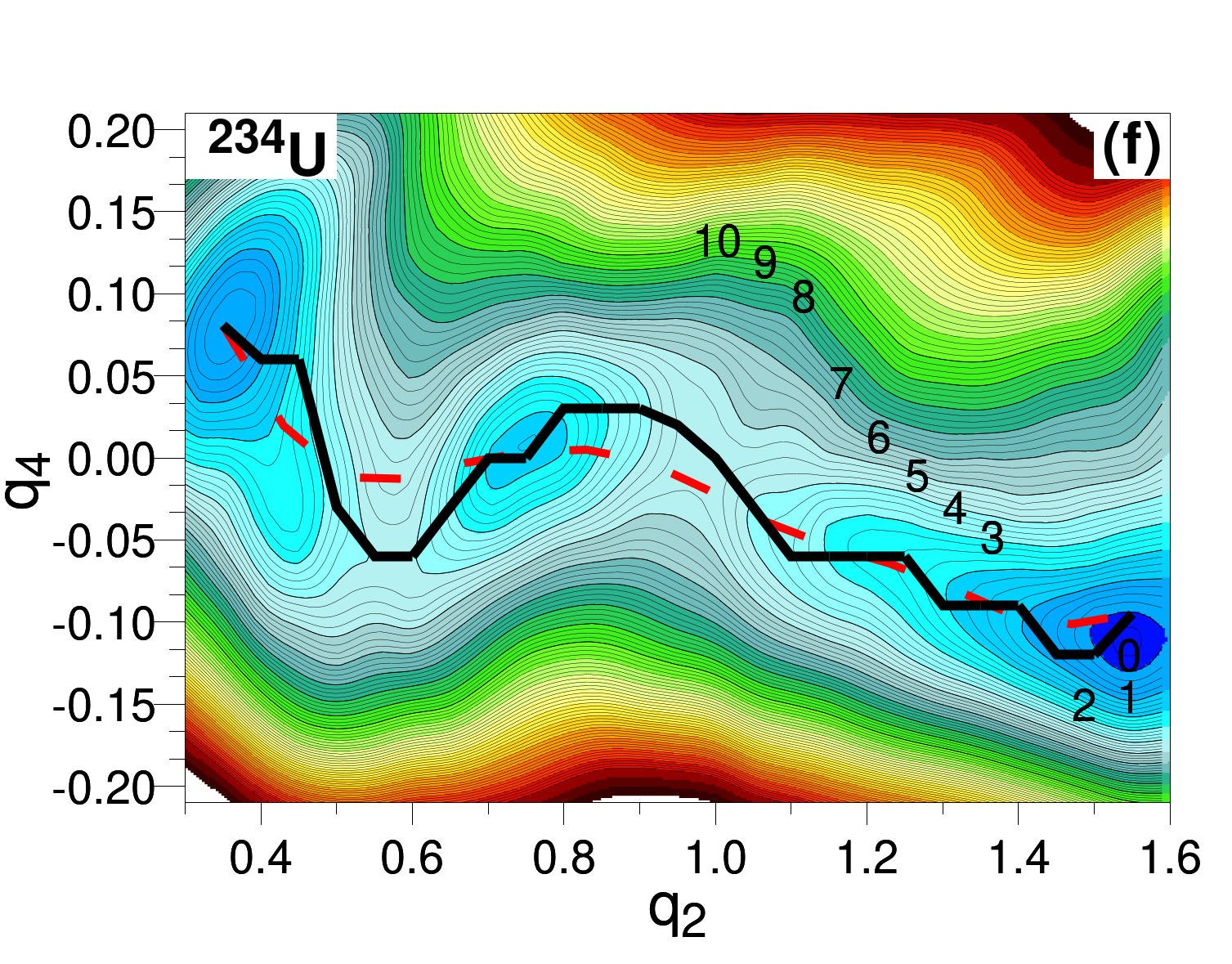}\\
      \includegraphics[scale=0.15]{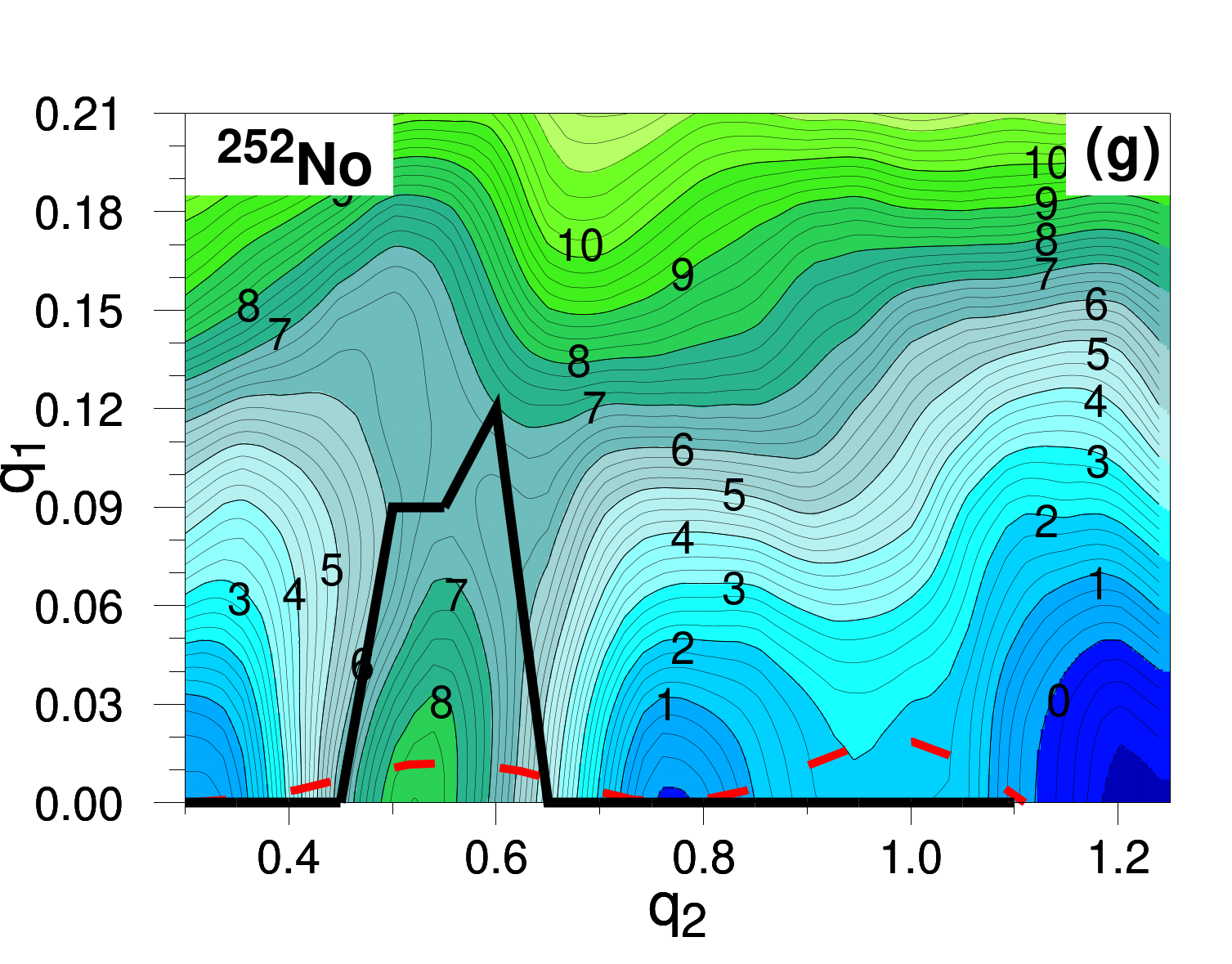} 
      \includegraphics[scale=0.15]{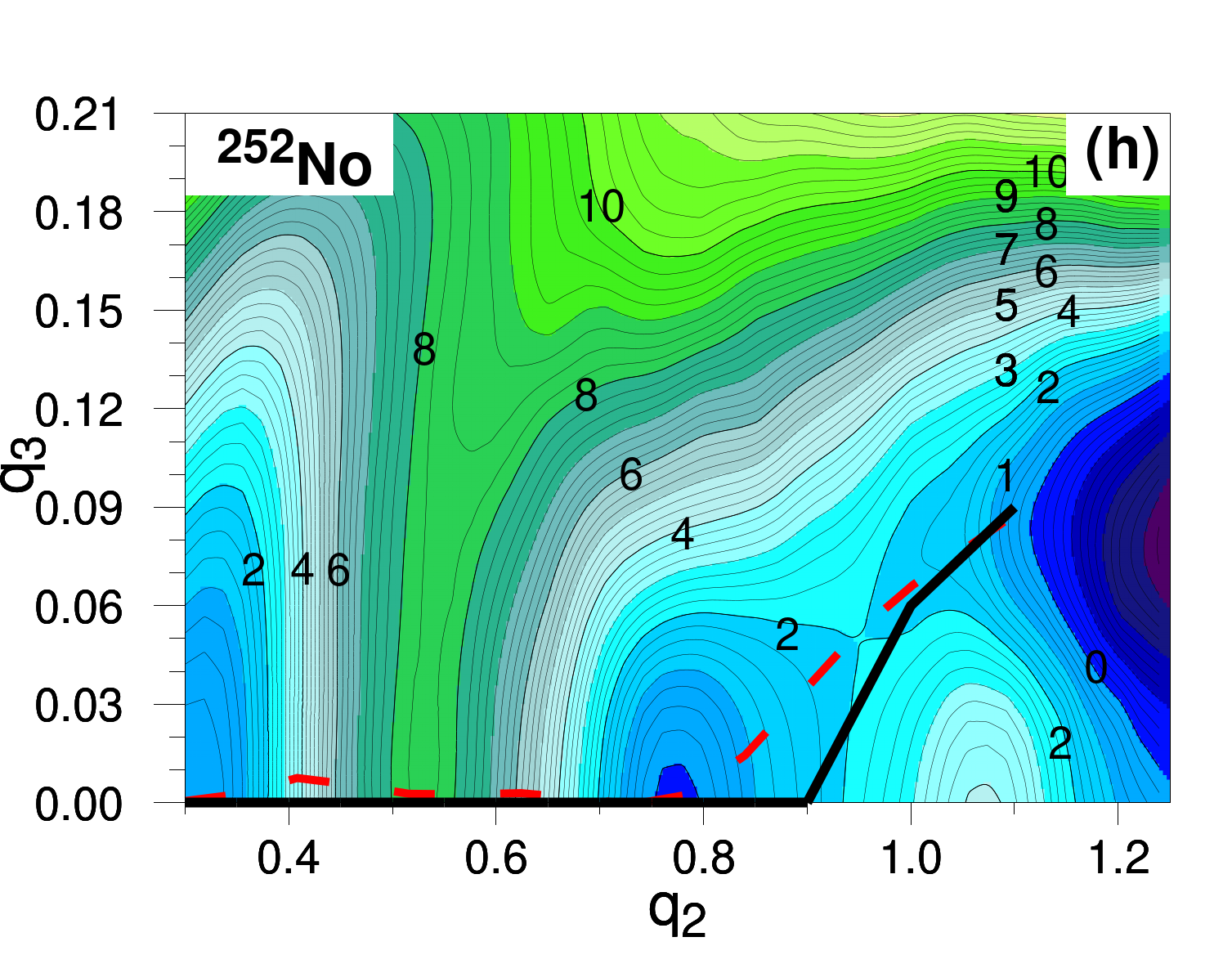}\\ 
\end{figure*}

\begin{figure*}[ht!]
      \includegraphics[scale=0.15]{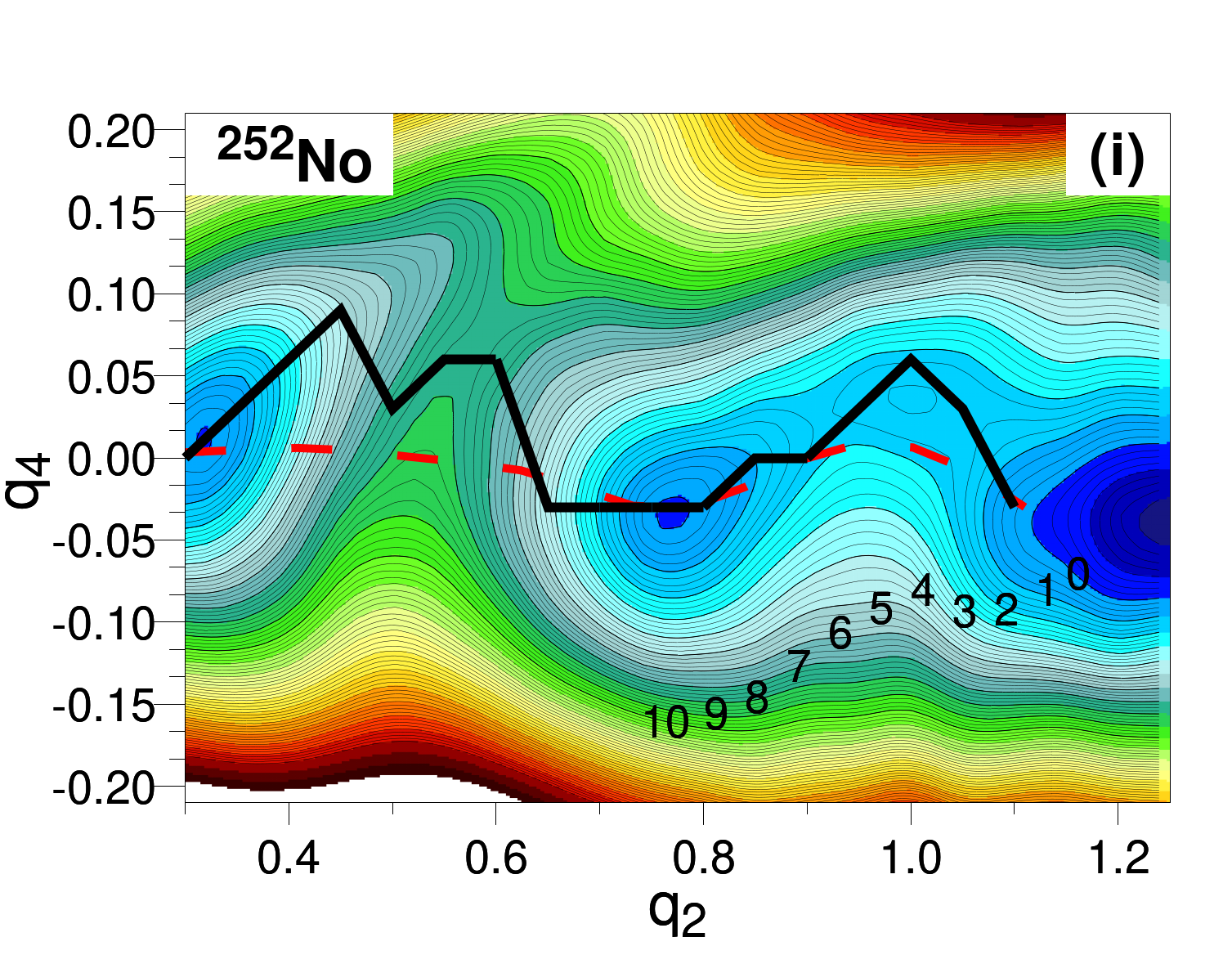}
\caption{Potential energy surfaces for $^{230}$U (a-c), $^{234}$U (d-f) and $^{252}$No (g-i) isotopes projected onto the $(q_2,q_1)$, $(q_2,q_3)$ and $(q_2,q_4)$ deformation subspaces. The projection is performed in such a way that the other two variables $q_k(q_2)$ and $q_{k'}(q_2)$ take values that minimize the action (\ref{LAP}) between the equilibrium and the exit points.  The dashed red and solid black curves correspond respectively to the LAP obtained with the hydrodynamical mass tensor and the least-energy path (LEP).}
\label{LAP_plots}
\end{figure*}


Please notice that the magnitude of the center-of-mass distance $R_{12}$ depends essentially on the elongation $q_2$ and only weakly on the left-right asymmetry and the neck formation parameters, $q_3$ and $q_4$, respectively. For that reason the least-action fission path obtained with the phenomenological mass of Eq.\ (\ref{masa}) cannot be called fully dynamical.

According to the main concept, the parameter $k$ in (\ref{masa}) is chosen so as to ensure that the value of $B(R_{12})$ along the fission barrier (in the vicinity of $q_2 \approx 1$) is close to the value of the hydrodynamical mass tensor in that area. 
\\[-2.0ex]

 At the same time, it should reproduce the asymptotic behaviour of the rigid-body inertia when a nucleus splits into two fragments, in which case the inertia of the strongly elongated nucleus, close to the scission configuration, should smoothly merge into the reduced mass of the two fragments. It turned out that the optimal value of this parameter is $k=11.5$.
\\[-2.0ex]

%
%

Let us now explain an efficient method to determine the least-action path in our 4D deformation space with the full hydrodynamical mass tensor, which will then be used to calculate the tunneling probability through the fission barrier to determine the spontaneous fission half-lives, a method based on the concept first introduced by Ritz \cite{ritz} and successfully used 
in particular to study the spontaneous fission process (see e.g.\ Refs.\ \cite{baranritz,baran81,LS99}.

In order to define any path in this deformation space, one first of all notices that any continuous and bounded function over a given finite interval of its arguments 
can always be approximated by a Fourier type expansion, involving only $sin$ functions on top of an {\it average path} whenever the endpoints of that path are fixed. 
 In our case these endpoints are the ground state and the exit point which is characterized by the same energy as the ground state. Defining that average path under the barrier by a straight line 
in the 4D deformation space 
connecting the ground state and the chosen exit point and considering the elongation parameter $q_2$ as the essential variable responsable for the fission process, one can always approximate the deformation parameters $q_1$, $q_3$ and $q_4$ along the least-action path as functions of $q_2$ in the following way:

\begin{eqnarray}
  q_{\nu}^{({\rm LAP})}(q_{2})= \left[ q_{{\nu}_{g.s.}} + \frac{(q_{{\nu}_{exit}} - q_{{\nu}_{g.s.}})(q_{2} - q_{{2}_{g.s.}})}{q_{{2}_{exit}} - q_{{2}_{g.s.}}} \right. \;\;\;\;
\nonumber\\
\hspace{-0.7cm}  
   + \left. \sum_{\ell=1}^{N_F}a_{\ell}\,sin\left(\ell\pi\frac{q_{2} - q_{{2}_{g.s.}}}{q_{{2}_{exit}} - q_{{2}_{g.s.}}} \right) \right], \;\; \nu=1,3,4 \;\;\;\;
\label{path}\end{eqnarray}
where the amplitudes $a_{\ell}$ of the series expansion are treated as variational parameters relative to which the minimum of the action integral (\ref{LAP}) is being searched. 
The upper limit $N_F$ of the Fourier series expansion in each direction of the least-action path (\ref{path}) has to be chosen such that the final result for the tunneling probability becomes essentially independent of $N_F$. 
We have found that a value of $N_F=8$ turns out to be sufficient to obtain a very good convergence of the Fourier series and thus a well converged tunneling probability.

Having found the least-action integral value (with respect to the $a_{\ell}$ amplitudes), one thus obtains the evolution of this path for a given nucleus in the considered 4D deformation space.   
\section{Results}   

In Figs.~\ref{LAP_plots} we present, for $^{230}$U, $^{234}$U and $^{252}$No, the projections of the full 4D PES onto the 2D sub-spaces $(q_2,q_1)$, $(q_2,q_3)$ and $(q_2,q_4)$, where the other two deformation parameters are the functions of the elongation $q_2$ that minimize the action integral (\ref{LAP}) between the ground state and the true exit point. The evolution of the LAP obtained when using the above discussed hydrodynamical mass tensor in these landscapes is indicated by the dashed red line. These isotopes have been chosen to cover the region from light to heavy actinides. As can be seen, the PES and the associated LAP in these extreme cases have different characteristics. 

In the lighter actinides, due to the importance of the shell effects, the PES is showing a stronger deformation dependence than in the heavy No isotope. Consequently, the fission barriers in uranium isotopes, unlike in nobelium, are generally higher and longer before reaching the scission configuration. Already from this quick qualitative analysis, one can expect a shorter half-life for nobelium as compared to uranium, an analysis which turns out to be consistent with the experiment.
As can already be concluded from Eq.~(\ref{LAP}), the final course of the LAP in the multi-dimensional deformation space is dictated by the interplay between the deformation dependent PES and the inertia tensor. This is one of the reasons why the LAP is usually shorter than the least-energy path (LEP) and passes, in general, through higher energy configurations (sometimes by as much as 2 MeV) as compared to the corresponding LEP. The actions along both these trajectories can therefore differ significantly, thus causing sometimes a difference of several order of magnitude in the estimates of the fission half-lives.

%

As one can see from part (e) of Fig.~\ref{LAP_plots} the LAP for $^{234}$U starts from the mass-symmetric ground state, stays left-right symmetric ($q_3=0$) up to the second minimum and then evolves towards asymmetric shapes around $q_2 \approx 0.7$, leading finally to an asymmetric fission valley at a value of $q_3 \approx 0.08$. One thus concludes that beyond the second minimum it is absolutely crucial to take into account the mass asymmetry degree of freedom. One notices that for the $^{234}$U nucleus the LEP and the LAP stay fairly close to each other in the $(q_2,q_3)$ plane. When looking at the $(q_2,q_4)$ subspace for the same nucleus (see panel (f) of Fig.~\ref{LAP_plots}), one finds that the LAP shows only a small deviations from a linear behaviour between a compact ground-state shape at $q_4 \approx 0.08$ and a medium-elongated necked-in shape at $q_4 \approx -0.12$.
Regarding the non-axiality in panel (d), the LAP between the ground state and the second minimum goes through rather moderate non-axial shapes and returns to the axial path at $q_2=0.65$.
A similar trend is observed in the super-heavy $^{252}$No nucleus, panels (g)-(i). However in this case the LAP ends around $q_2=1.1$, and is thus much shorter than in $^{234}$U.  For $^{230}$U (panels (a)-(c)), on the contrary, the $q_3$ deformation is almost negligible in the initial and final stages of the LAP
whereas at intermediate elongations it reaches values beyond $q_3\approx 0.1$ to bypass the energy maximum (barrier) peaked at $q_2\approx 1$. In both the discussed uranium isotopes their LAPs $q_4(q_2)$ presented in panels (c) and (f) have comparable shapes. Interestingly, the LAP in this nucleus prefers fully axial shapes throughout its course, even though the LEP passes through energetically slightly lower, non-axial configurations at very small deformations just beyond the equilibrium point. In order to keep the computation time within reasonable limits, without making any compromise on the precision of our results, we are able to consider up to $N_F^{tot}=3\times N_F=24$ harmonic components of the Fourier series in Eq.\ (\ref{path}).
In such a large number of dimensions, one may clearly encounter a problem of distinguishing between some local and the global minimum of the action integral. To avoid this behaviour, we start the calculations for each nucleus with 
a low value of $N_F$, e.g. $N_F=3$  for each of the three functions $q_1(q_2)$, $q_3(q_2)$ and $q_4(q_2)$, on top of the average path. The value of $N_F$ is then gradually increased, checking after each step whether convergence is obtained. 
It turns out that restricting ourselves to the first few components of these series, like e.g.\ $N_F=6-8$ leads to LAPs in Fig.\ \ref{LAP_plots} that visually cannot be distinguished from the ones obtained with larger values of $N_F$.  
\\[-2.0ex]

Having calculated the values $S$ of the action, one can finally determine the spontaneous fission lifetime using the standard WKB relations \cite{SSS95}.
\begin{equation}
    T_{1/2}^{({\rm sf})}=\frac{2\pi\ln(2)}{\omega_0} \left(1+e^{2S}\right),
\end{equation}
where $E_{\rm {ZPE}} \!\approx\! \frac{1}{2}\hbar\omega_0$ stands for the zero-point vibration energy which is usually taken to be in the range of $0.5 - 1\;$MeV. In the present work we have taken a value of $E_{\rm {ZPE}}=0.5\,$MeV.
%
%
%
%
%

\begin{figure}[htbp]
\begin{minipage}[t]{0.45\textwidth}
\includegraphics[scale=0.4]{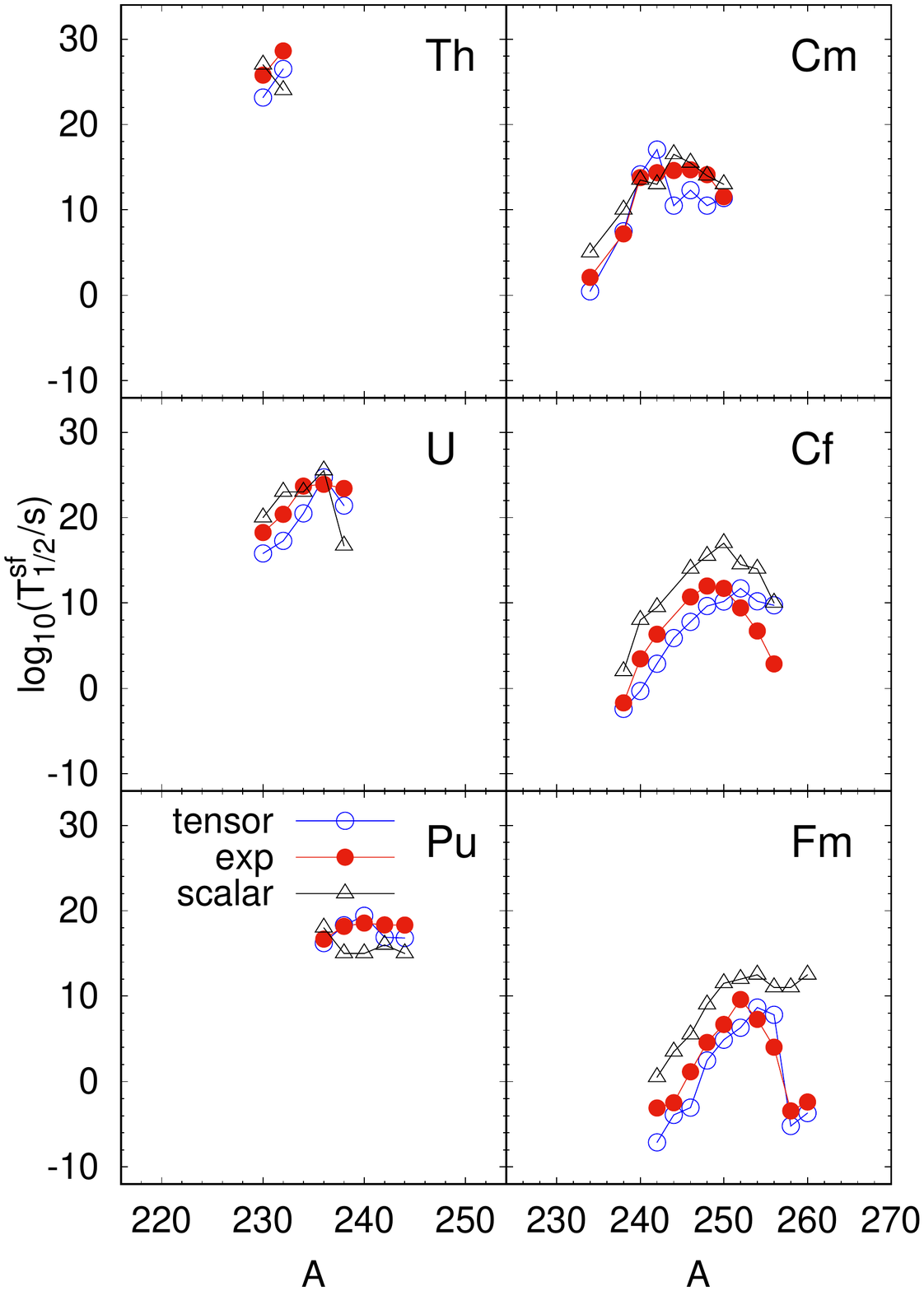}
\end{minipage}
\hfill
\begin{minipage}[t]{0.45\textwidth}
\includegraphics[scale=0.4]{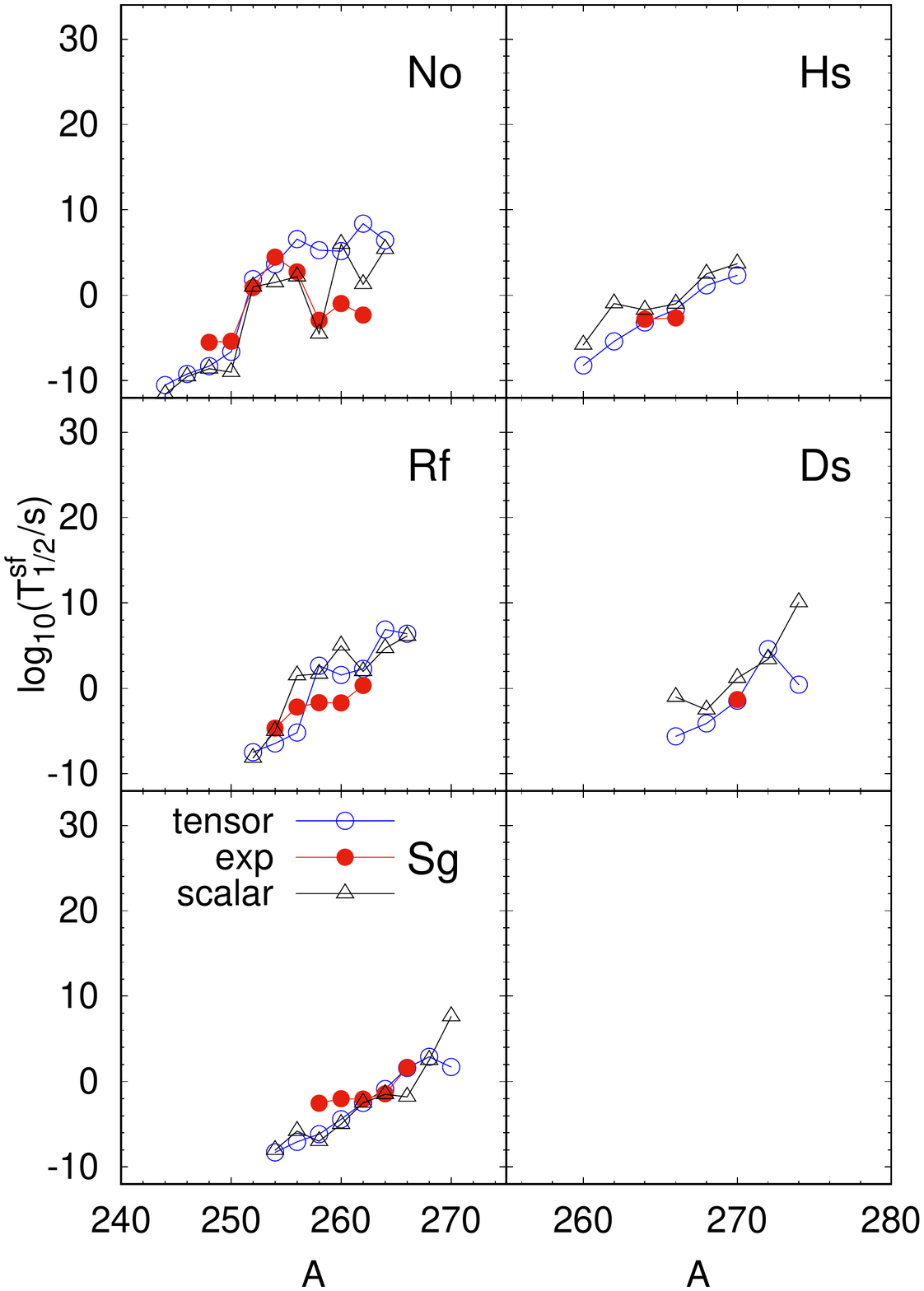}
\end{minipage}
\hfill

\caption{(Top panel) Comparison of spontaneous fission half-lives for actinide nuclei obtained in our 4D WKB approach with the irrotational flow hydrodynamical mass tensor (open circles) and the phenomenological inertia with 3D PES, Eq.\ (\ref{masa}) (open triangles) with the experimental data (full red circles). (Bottom panel) Same but for super-heavy elements from nobelium to darmstadtium.}
\label{3D_WKB}
\end{figure}

Spontaneous fission half-lives are determined for selected isotopes of some actinide nuclei, namely thorium (Th), uranium (U), plutonium (Pu), curium (Cm), califormium (Cf), and fermium (Fm) and for super heavy isotopes of nobelium (Nb), rutherfordium (Rf), seaborgium (Sg), hassium (Hs) and darmstadtium (Ds) for which experimental data are available \cite{NUDAT}. 
The results of the calculations obtained with the 
rescaled (adjusting the value of $\beta$) 
hydrodynamical mass tensor and the phenomenological mass formula (\ref{masa}) are presented in Fig.~\ref{3D_WKB} together with the measured values. The data for the different isotopes calculated within the above presented approach are given as open blue circles and black triangles, while the experimental data are shown as full red circles.
%
%
%
%
%

In order to obtain some
systematics for the spontaneous fission half-lives displayed in Fig.~\ref{3D_WKB} for all isotopic chains of actinides and super-heavy elements up to $Z\!=\!110$,  we have 
adjusted the parameter $\beta$ in Eq.~(\ref{LAP}) through a $\chi^2$ minimization to all the 39 measured half-lives of actinide nuclei from thorium to fermium presented in this figure. 
It is found that the half-lives in all presented actinides are well reproduced with a value of $\beta\!=\!5$. 
By choosing a smaller sample of nuclei we have made sure that the value of $\beta \!=\!5$ is, indeed, practically independent of the chosen sample. Such a $\beta$ value 
ensures that the logarithm of the evaluated half-lives in super-heavy nuclei stays within reasonable limits of approximately $2-3$  
(in $1/s$ units) which is comparable with other recent evaluations \cite{wardaPhysRevC.86.014322,pomshe}. Let us notice at this point that, as shown e.g.\ in Ref.~\cite{PomGozdz}, the hydrodynamical inertia used in our approach may differ from the commonly used microscopic mass tensors obtained within the cranking or the GCM+GOA model by almost a factor of $3-5$ on average, particularly for compact nuclear shapes, ($q_2 \approx 1$) where the fission barrier is located.
Once the $\beta$ value has been fixed, the spontaneous fission half-lives calculations are performed for super-heavy elements and compared with the experimental data. 
\\[-2.0ex]

Let us mention in this connection that the hydrodynamical inertia tensor has been successfully used in calculations of fission properties determined by shapes lying close to the scission configuration, such as fragment mass or charge distributions, whereas the barrier penetration occurs at significantly lower elongations around the fission barrier (see e.g. Ref.~\cite{chipom}). 
As seen in Fig.~\ref{Bij_maps}, the inertia components $B_{22}$ and $B_{33}$, most crucial for the barrier penetration, are much smaller in the vicinity of the barrier region than the ones close to the scission point. One has to keep in mind, however, that the pure hydrodynamical approach seems, in its original form, not really well suited
for a reliable description of the effective inertia near the barrier. 
The phenomenological mass parameter (\ref{masa}), on the other hand, contains the rigid-body inertia as the essential contribution together with a term (controlled by the parameter $k$) determined by the difference between the rigid body and the irrotational flow inertia, thus making it more reliable in the fission-barrier region.
Investigating through the results presented in Fig.~\ref{3D_WKB} the capacity of our approach using the hydrodynamical inertia tensor to reproduce the experimental fission half-lives for actinides, one could be quite satisfied. 
There are, however, a few cases, that stick out from their isotopic systematics by several orders of magnitude, which we would like to understand a little better. These are e.g.\ some isotopes of Cm and the heaviest Cf nuclei.
To explain these discrepancies one may refer e.g.\ to a recent work \cite{KP22}, where it is shown, within a simple analytical 1D WKB approach, similar to the so called \'Swiatecki-like systematics \cite{swiatecki55} of spontaneous fission half-lives, that the main quantity determining the fission half-life is the fission barrier height $E_B$. Its dependence on other properties of the fission barrier, including the barrier width, is already somehow absorbed in the adjustable function $f(E_B)$ (given by equations (24) and (25) of Ref.~\cite{KP22}), common for all heavy and super-heavy elements. 
At this point it may be worth to recall that even a small change in the fission barrier height, leads to a substantial decrease or increase of the tuneling probability and, as a consequence, produces longer or shorter fission half-lives.
Another interesting case is the one of the $^{232-234}$U isotopes, where our macroscopic-microscopic model underestimates the fission barrier heights by about $1-2$ MeV (see also Ref.~\cite{Josepaper}) causing an underestimation of the resulting fission half-lives by as much as some 2-4 orders of magnitude. A similar effect can also be observed for the $^{242-246}$Cm isotopes, where the discrepancies between the experimental and theoretical first and second barriers are the largest throughout the whole isotopic chain.
\\[ -2.0ex]

The reason for the overestimation of the lifetimes for the superheavy $^{258-262}$No and $^{256-260}$Rf isotopes is presumably similar to that for the aforementioned actinides except that we do not know yet the experimental barriers to be able to make quantitative comparisons.

It is interesting to also present in Fig.~\ref{3D_WKB} the results of half-lives calculations for actinide nuclei obtained with the phenomenological inertia of Eq.\ (\ref{masa}) at the three-dimensional PES, where the total 4D energy function is minimized, for a given $(q_2,q_3,q_4)$ point, with respect to the nonaxiality parameter $q_1$. 
Comparing the results obtained with both here presented approximations, one observes that the half-lives with the 4D WKB dynamics and full hydrodynamical inertia tensor with a scalling factor $\beta\!=\!5$ (common for all tensor components) are in a majority of cases much closer to the experimental data than those obtained with the phenomenological scalar mass given by Eq.\ (\ref{masa}). This shows, we believe, that considering the full description of the inertial properties of a system at each deformation point, as well as including as many degrees of freedom relevant for the fission dynamics as ever possible leads to a substantial improvement of the fission-lifetime estimates.

\section{Conclusions}

Spontaneous fission half-lives for nuclei in the range $90 \!\leq\! Z \!\leq\! 104$ have been determined in the macroscopic-microscopic approach together with the Lublin-Strasbourg Drop model, a mean-field generated by a Yukawa-folding procedure and a constant $G$ seniority BCS pairing treatment with a GCM+GOA particle-number projection.
The dynamics of the fission process have been simulated by the semiclassical WKB method with the least-action integral describing the evolution of the nucleus in a four-dimensional deformation space given by the expansion coefficients of a Fourier shape parametrization which stand for elongation, mass asymmetry, non-axiality and neck degrees of freedom.
\\[ -2.0ex]

In order to take into account the variation of the collective inertia along the fission path, we have used in the action-integral expression the irrotational flow mass tensor which has been scaled by a factor of 5
in order to reproduce fission half-lives in the actinide region.
Since the resulting least-action path to fission tends, to some extent, to omit states where the inertia changes dramatically, due to the presence of shell effects, 
the use of this effectively macroscopic model of collective inertia should reduce the numerical instabilities in the search of the minimum of the action integral.
For a comparison, we have also performed similar calculations of fission lifetimes with a collective mass parameter, Eq.\ (\ref{masa}), which has also led to a quite reasonable reproduction of fission half lives over a wide range of actinide and super-heavy nuclei.

One notices that both these inertia approaches generally yield quite close values of the spontaneous fission half-lives $T_{1/2}$, particularly in super-heavy nuclei and in the actinide isotopes of thorium, uranium, plutonium and curium, while in californium and fermium, the use of the collective mass parameter, Eq.\ (\ref{masa}) and the 3D PES, leads to a mean deviation reaching several orders of magnitude relative to the experimental results. 
One should also be aware of the fact that the phenomenological mass formula (\ref{masa}) is a kind of hybrid approach which, in order to reproduce the spontaneous fission half-lives, combines two different evaluations of the nuclear collective inertia, namely the hydrodynamical and the rigid-body approaches, and   requires, to do that, 3 adjustable parameters, whereas our approach using the irrotational flow mass tensor yields better results with a single adjustable parameter, namely the parameter $\beta$ in the expression (\ref{LAP}) for the action integral.
Let us mention at this point that no zero-point energy correction is needed in our approach and would, in fact, be in disaccord with the philosophy of our macroscopic-microscopic model which, similar to a mean-field Hartree-Fock type framework, describes the nuclear energy by some kind of variational approach, where there is no place for an artificial increase or lowering of the ground-state energy.

One should also keep in mind that spontaneous-fission is only one of several possible nuclear decay channels, competing with the emission of light particles (like {\it n} or {\it p}), $\gamma$ quanta, or the emission of light clusters (like e.g. $\alpha$ particles). The competition between fission and these other processes is something we are presently working on, and will be the subject of a forthcoming publication.



\begin{acknowledgements}
This work is supported by the COPIN-IN2P3 agreement (Project No. 08-131)
between the Polish and French nuclear laboratories and the Polish National
Science Center (Project No. 2018/30/Q/ST2/00185).
The research work of A.Z.\ is part of the project No. 2021/43/P/ST2/03036 co-funded by the National Science Centre and the European Union Framework Programme for Research and Innovation Horizon 2020 under the Marie Skłodowska-Curie grant agreement no.\ 945339. For the purpose of Open Access, the author has applied a CC-BY public copyright licence to any Author Accepted Manuscript (AAM) version arising from this submission.
                                                                                                                                                                                                                                                                                                                                                                                                                           
\end{acknowledgements}

\bibliographystyle{apsrev}
\bibliography{references}

\end{document}